\documentclass[lettersize,journal]{IEEEtran}
\usepackage{amsmath,amsfonts,amssymb,amsthm}
\usepackage{algorithmic}
\usepackage{algorithm}
\usepackage{array}
\usepackage[caption=false,font=footnotesize,labelfont=rm,textfont=rm]{subfig}
\usepackage{textcomp}
\usepackage{stfloats}
\usepackage{url}
\usepackage{verbatim}
\usepackage{graphicx}
\usepackage{cite}
\usepackage{xcolor}
\usepackage{booktabs}
\usepackage{multirow}
\usepackage{capt-of}
\usepackage{makecell} 
\begin{document}

\title{Pinching-Antenna System With Movable Waveguides: Modeling and Optimization}

\author{Jingze Ding, \IEEEmembership{Graduate Student Member, IEEE},
		Zijian Zhou, \IEEEmembership{Member, IEEE}, 
		Bingli Jiao, \IEEEmembership{Senior Member, IEEE},
		and Rui Zhang, \IEEEmembership{Fellow, IEEE}
\thanks{This work is supported in part by National University of Singapore under Research Grants A-8003646-00-00 and A-8003676-00-00, and in part by China Scholarship Council. An earlier version of this paper was presented in part at the IEEE WCNC 2026 \cite{conf}. \textit{(Corresponding author: Rui Zhang.)}}
\thanks{Jingze Ding is with the School of Electronics, Peking University, Beijing 100871, China, and also with the Department of Electrical and Computer Engineering, National University of Singapore, Singapore 117583 (e-mail: djz@stu.pku.edu.cn).}
\thanks{Zijian Zhou is with the School of Science and Engineering, The Chinese University of Hong Kong, Shenzhen 518172, China (e-mail: zijianzhou@link.cuhk.edu.cn).}
\thanks{Bingli Jiao is with the School of Computing and Artificial Intelligence, Fuyao University of Science and Technology, Fuzhou 350109, China. and also with the School of Electronics, Peking University, Beijing 100871, China (e-mail: jiaobl@pku.edu.cn).}
\thanks{Rui Zhang is with the Department of Electrical and Computer Engineering, National University of Singapore, Singapore 117583 (e-mail: elezhang@nus.edu.sg).}
}
\maketitle

\begin{abstract}
This paper proposes a movable waveguide (MW)-enabled pinching-antenna system (PASS), in which each waveguide is connected via a flexible cable and can be linearly moved by drivers. By simultaneously moving the MWs and the pinching antennas (PAs) on them, MW-enabled PASS can effectively track user locations and form flexible array geometries for efficient beamforming. We first examine the special case with a single user and derive the closed-form solutions for the optimal MW positions as well as an upper bound on the user rate. Furthermore, we develop a two-step optimization algorithm to maximize the achievable rate for the user, where the first step determines the optimal MW positions using the derived closed-form solutions, and the second step alternately optimizes the PA positions through a one-dimensional (1D) local search based on the user location. Then, for the general multi-user scenario, we derive the upper bounds on the minimum rate among all users. To maximize their minimum rate, we propose a low-complexity two-scale optimization algorithm, where the large-scale global search coarsely determines the MW and PA positions, followed by a small-scale local search to finely tune them. In addition, a two-timescale optimization scheme based on statistical channel information is investigated to reduce the mechanical movement overhead of the MWs. Simulation results demonstrate that the proposed scheme achieves performance close to the derived bounds. It also flexibly adapts to different user distributions compared with the conventional PASS employing dense or sparse fixed-position waveguides (FPWs), as well as fixed-position antenna (FPA) schemes.
\end{abstract}
\section{Introduction}
As a cornerstone of modern wireless communications, multiple-input multiple-output (MIMO) technology has played a pivotal role in improving the spectral efficiency of past and present wireless systems \cite{MIMO2,MIMO3}. To address the explosive growth in demand expected in future wireless networks, a new class of flexible MIMO systems has recently emerged, including movable/fluid antennas \cite{MA1,MA2,MA3,MA4} and polarization-reconfigurable antennas \cite{PRA1,PRA2,PRA3,PRA4}. These technologies introduce new degrees of freedom (DoFs) to improve wireless channels, allowing wireless transceivers to more efficiently allocate spatial resources according to diverse application requirements. Although these technologies have been shown to provide promising performance improvements, they are mainly employed to mitigate small-scale channel fading but are still limited in addressing large-scale path loss. Furthermore, these systems generally lack hardware flexibility, as once they are deployed, adding or removing antennas is practically difficult and inconvenient.
\subsection{Pinching-Antenna System}
To address the aforementioned limitations, the authors of \cite{PASS1} recently proposed the pinching-antenna system (PASS) as a new flexible antenna architecture. PASS employs the dielectric waveguide as the transmission medium, on which small dielectric particles are attached at designated positions to radiate electromagnetic waves propagating within the waveguide \cite{PASS2}. These dielectric particles are typically mounted on the tips of plastic clips, referred to as pinching antennas (PAs), which can be conveniently attached/detached to/from the waveguide or even dynamically moved along it, thereby enabling dynamic array reconfiguration \cite{PASS3}. Unlike other flexible antenna systems, the waveguides in PASS can be extended to arbitrary lengths, thus allowing antennas to be deployed in close proximity to users. This capability facilitates the establishment of strong and stable line-of-sight (LoS) channels, thereby effectively reducing large-scale path loss. Moreover, by simply adding or removing dielectric particles, PASS requires only low-cost implementation. 

Due to the above advantages, the applications of PASS in wireless communication and sensing systems have recently attracted increasing research attention. In \cite{PASS4}, the authors conducted the first theoretical analysis of PASS with both single and multiple waveguides, and highlighted that the impact of user/waveguide deployment on system performance is an important direction for future research. In \cite{PASS5}, a hardware model of PASS grounded in physical principles was developed, in which the PA was modeled as an open-ended directional coupler and its electromagnetic characteristics were analyzed using coupled-mode theory. Moreover, the authors of \cite{PASS6} proposed a novel adjustable power radiation model for PASS, where the power radiation ratio of each PA can be flexibly controlled by tuning the spacing between the PAs and the waveguides. Furthermore, the authors of \cite{PASS7} proposed the centralized PA deployment to exploit beamforming gain and the distributed PA deployment to harness multiplexing gain for serving multiple users. To determine the optimal PA positions, the authors of \cite{PASS8} investigated optimization-based and learning-based methods for solving the antenna position optimization problem. Besides, an element-wise optimization framework was proposed in \cite{PASS9} to sequentially optimize the position of each PA using a low-complexity one-dimensional (1D) search, thereby addressing the sum-rate maximization problem in multi-user communications. In addition, the applications of PASS in integrated sensing and communications (ISAC) \cite{PASS_ISAC1,PASS_ISAC2,PASS_ISAC3,PASS_ISAC4}, non-orthogonal multiple access (NOMA) \cite{PASS_NOMA1,PASS_NOMA2,PASS_NOMA3,PASS_NOMA4}, and physical layer security (PLS) \cite{PASS_PLS1,PASS_PLS2,PASS_PLS3,PASS_PLS4} have also been explored.
\subsection{Motivations and Contributions} 
While a substantial number of studies have investigated PASS, most have focused on resource allocation in various scenarios, yet leaving the following key challenges unaddressed.
\begin{itemize}
	\item \textit{Uniform and Sparse Waveguide Deployment}: Existing works on PASS primarily assume that the waveguides are uniformly spaced across the entire coverage area, which leads to relatively large inter-waveguide spacing. Nevertheless, this type of deployment is not effective for supporting users located in arbitrary positions. In particular, only the PAs on nearby waveguides can effectively serve the users, whereas those on distant waveguides are ineffective. At present, the impact of different waveguide deployment strategies, such as non-uniform and/or small inter-waveguide spacing, on PASS performance has remained largely unexplored.
	\item \textit{Fixed-Position Waveguide Configuration}: The majority of existing PASS research assumes the fixed-position waveguide (FPW), which limits the reconfigurability of antenna positions, since PAs are constrained to move only along the line segments specified by the FPWs. As a result, PASS with FPWs cannot fully adapt to varying user distributions. For example, a horizontally deployed waveguide is unable to resolve users distributed along the vertical direction, regardless of the number of PAs mounted on it. Therefore, it is essential to investigate more flexible PASS designs that can achieve robust performance under diverse user distribution scenarios.
\end{itemize}

The performance of PASS depends critically on the PA positions. From an information-theoretic perspective, for a given radio environment (e.g., a user distribution), there exists an optimal set of PA positions that can attain the maximum achievable capacity. However, the feasible PA positions are fundamentally constrained by the waveguide deployment, since each PA can only be positioned along its waveguide. To unlock the full potential of PASS, we propose in this paper a new movable waveguide (MW)-enabled PASS model, where each MW is connected to the radio frequency (RF) chain through a flexible cable. By mounting the MWs on a mechanical slide and attaching them to dedicated motors \cite{EE}, they can move linearly (horizontally or vertically) within a predefined region\footnote{Unlike movable-antenna systems where antennas move within a fixed small-scale (in the order of several to tens of carrier wavelengths) feasible region, the large-scale MW movement in the proposed system can more significantly reconfigure the feasible region for antenna repositioning, which introduces a fundamentally different DoF from existing movable-antenna and PASS models.}, as illustrated in Fig. \ref{sysmodel}. The proposed MW-enabled PASS addresses the limitations of conventional PASS in two aspects. First, the MWs can adjust the inter-waveguide spacing to achieve non-uniform waveguide placement. In addition, they can be moved to locations close to the users based on their positions, thereby ensuring that each MW is effectively utilized. Second, the MWs together with the PAs movable along them provide a two-dimensional (2D) region for all PAs to move freely within, thereby enabling the joint optimization of the MW and PA positions to flexibly and dynamically reshape the antenna array geometry to serve users with different spatial distributions. To demonstrate the above benefits, we consider the MW-enabled PASS for multi-user downlink communications. The main contributions of this paper are summarized as follows:
\begin{figure*}[!t]
	\centering
	\includegraphics[width=0.8\linewidth]{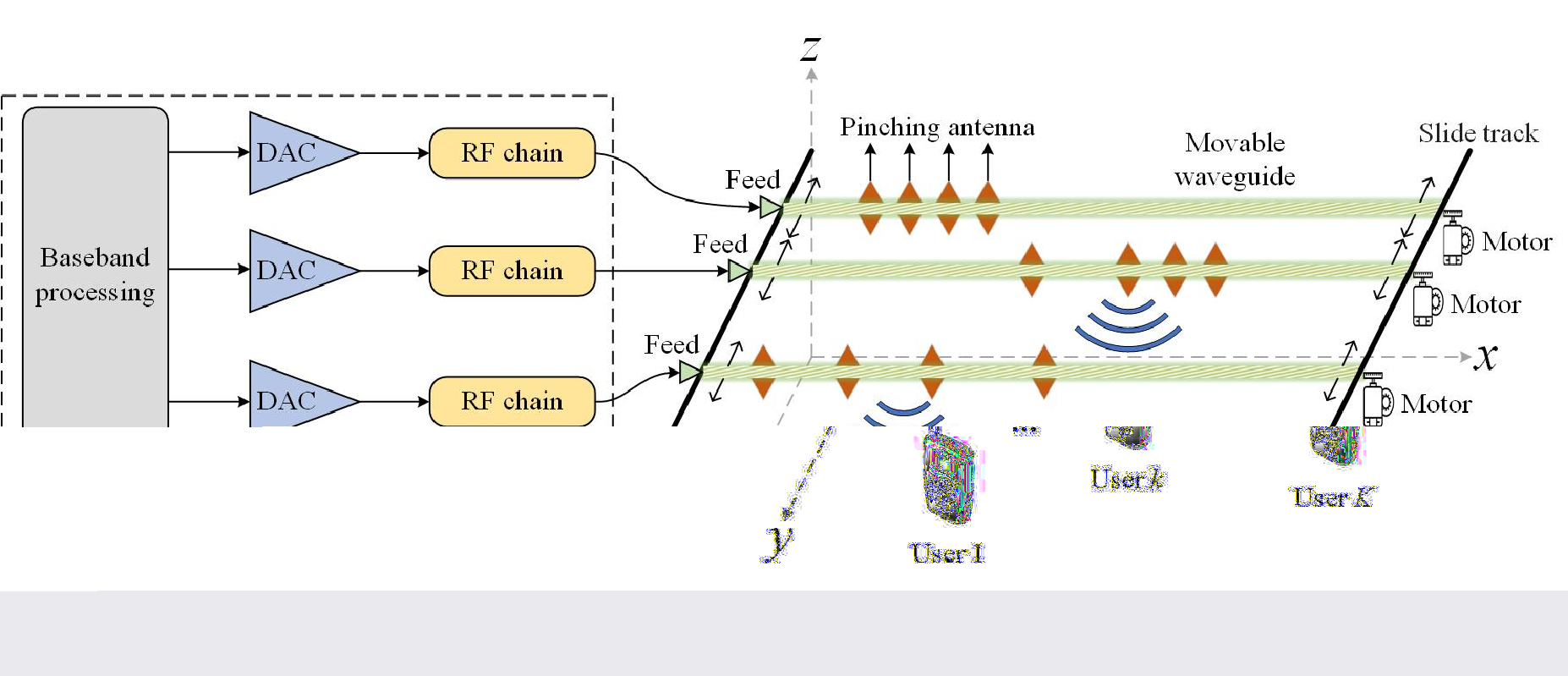}
	\caption{Illustration of the MW-enabled PASS for downlink communication.}
	\label{sysmodel}
\end{figure*}
\begin{itemize}
	\item First, we propose the MW architecture and establish the channel and signal models for MW-enabled PASS in multi-user downlink communications. To guarantee fairness in quality of service (QoS) for each user, we consider the users' maximized minimum (max-min) rate as the performance metric, which is achieved by maximizing the minimum rate among all users by jointly optimizing the MW and PA positions, as well as the transmit beamforming vectors, subject to the movement-related constraints and the maximum transmit power constraint.
	\item Next, for the special case of a single-user system, we analyze both the scenarios of single-PA and multiple-PA deployment per MW and derive the closed-form solutions for the optimal MW positions. Furthermore, an upper bound on the user rate is derived, based on which a low-complexity PA position optimization algorithm is developed to approach this performance bound.
	\item Then, for the general multi-user case, we derive upper bounds on the users' max-min rate and analyze the conditions under which these bounds can be achieved. In addition, to jointly optimize the MW and PA positions with low complexity, we propose a two-scale optimization algorithm, where a large-scale search determines the coarse MW and PA positions, followed by a small-scale search to refine them. Moreover, a two-timescale optimization scheme based on statistical channel information is proposed to reduce the MW movement overhead.
	\item Finally, we conduct extensive simulations to validate the advantages of the proposed MW-enabled PASS. The results demonstrate that PASS with MWs can track user locations to effectively reduce large-scale path loss and reconfigure the antenna array geometry to adapt to different user distributions. As a result, the proposed scheme achieves considerable performance gains over the conventional PASS with FPWs as well as fixed-position antenna (FPA) systems.
\end{itemize}

The rest of this paper is organized as follows. Section \ref{2} introduces the channel and signal models of the MW-enabled PASS and formulates the minimum-rate maximization problem. In Section \ref{3}, we first study a simple single-user scenario, which is then extended to the general multi-user case in Section \ref{4}. Next, simulation results and discussions are provided in Section \ref{5}. Finally, this paper is concluded in Section \ref{6}.

\textit{Notation:}  $a/A$, $\mathbf{a}$, $\mathbf{A}$, and $\mathcal{A}$ denote a scalar, a vector, a matrix, and a set, respectively. ${\left(  \cdot  \right)^{T}}$, ${\left(  \cdot  \right)^{H}}$, $\left|  \cdot  \right|$, and $\mathrm{tr}\left( \cdot\right) $ denote the transpose, conjugate transpose, absolute value, and trace, respectively. $\left\|\cdot\right\|_1$ and $\left\|\cdot\right\|_2$ denote the 1-norm and 2-norm of a vector, respectively. $\left\|\cdot\right\|_F$ represents the Frobenius norm of a matrix. $\angle \mathbf{x}$ denotes the phase of the complex vector $\mathbf{x}$. $(x)^+$ stands for $\max(x,0)$. $\mathbf{I}_K$ is the identity matrix of order $K$. $\mathcal{A} \backslash \mathcal{B}$ denotes the subtraction of set $\mathcal{B}$ from set $\mathcal{A}$. $\mathbb{Z}$, $\mathbb{R}$, and $\mathbb{C}$ denote the sets of integers, real numbers, and complex numbers, respectively. $\mathbb{R}^{M \times N}$ and $\mathbb{C}^{M \times N}$ represent the sets of real and complex matrices/vectors of dimension $M \times N$, respectively. $\mathbb{E}\left(\cdot \right) $ is the expectation of a random variable. $\mathcal{CN}\left( 0,\sigma^2 \right) $ represents the circularly symmetric complex Gaussian (CSCG) distribution with mean zero and variance $\sigma^2$. $\sim$ and $\triangleq$ stand for ``distributed as'' and ``defined as'', respectively.
\section{System Model and Problem Formulation}\label{2}
As shown in Fig. \ref{sysmodel}, we consider a multi-user downlink communication system enabled by a PASS equipped with MWs, and we adopt a global coordinate system to specify the positions of the PAs, MWs, and users. The base station (BS) is equipped with $N$ MWs, each aligned with the $x$-axis and capable of linear movement along the $y$-axis through the control of a motor attached to it, which jointly serve $K$ single-antenna users. There are $L$ PAs flexibly pinched along each MW\footnote{In practical deployment, multiple PAs can be pre-installed on the waveguides, and only $L$ of them are selectively activated to effectively achieve PA position adjustment.}, and thus the total number of PAs is $L \times N$. To support spatial multiplexing, each MW is connected to a dedicated RF chain through a flexible cable such that it can move freely. Both MWs and PAs are installed at a fixed height $h$. Let $S_\mathrm{x}$ and $S_\mathrm{y}$ denote the length and the allowable moving distance of each MW, respectively, which define a rectangular area of size $S_\mathrm{x} \times S_\mathrm{y}$. The three-dimensional (3D) Cartesian coordinate of the feed point for the $n$th ($1 \le n \le N$) MW is given by $\mathbf{u}_n^\mathrm{W} = {\left[ {0,y_n,{h}} \right]^T}$, where $y_n$, $0 \le y_n \le S_\mathrm{y}$, denotes the adjustable $y$-axis coordinate of this MW. We define $\mathbf{y}\triangleq \left[y_1, \ldots,y_N \right]^T\in \mathbb{R}^{N \times 1} $ as the set of all $y$-axis coordinates of the MWs, where $0 \le y_1 < y_2 <  \ldots  < y_N \le {S_\mathrm{y}}$. The position of the $l$th ($1 \le l \le L$) PA along the $n$th MW is ${\mathbf{u}_{n,l}} = {\left[ {{x_{n,l}},y_n,{h}} \right]^T}$, where $x_{n,l}$, $0 \le x_{n,l} \le S_\mathrm{x}$, is the adjustable pinched location of this PA along the $x$-axis. Let $\mathbf{X} \triangleq \left[ {{\mathbf{x}_1}, \ldots ,{\mathbf{x}_N}} \right] \in {\mathbb{R}^{L \times N}}$ denote all PA positions, where $\mathbf{x}_n \triangleq \left[x_{n,1}, \ldots, x_{n,L} \right]^T \in \mathbb{R}^{L \times 1}$ is the $x$-axis coordinates of PAs on the $n$th MW and satisfies $0 \le {x_{n,1}} < {x_{n,2}} <  \ldots  < {x_{n,L}} \le {S_\mathrm{x}}$. Furthermore, the location of the $k$th ($1 \le k \le K$) user is denoted by $\mathbf{u}_k^\mathrm{U} = {\left[ {x_k^\mathrm{U},y_k^\mathrm{U},z_k^\mathrm{U}} \right]^T}$, where we assume $0 \le x_k^\mathrm{U} \le S_\mathrm{x}$, $0 \le y_k^\mathrm{U} \le S_\mathrm{y}$, and $z_k^\mathrm{U}=0$ without loss of generality.
\subsection{Channel Model}\label{channel_model}
The PASS channel comprises two components: in-waveguide propagation and free-space propagation\footnote{This work considers a narrowband communication system, where the propagation delay variations across different PAs are assumed negligible relative to the symbol duration. For wideband systems, the larger bandwidth and the expanded array aperture of PASS may make such delay variations non-negligible, which may lead to inter-symbol interference (ISI). A practical solution is to adopt orthogonal frequency division multiplexing (OFDM)-based transmission with a sufficiently long cyclic prefix as a guard interval, which is an interesting topic for investigation in future work.}.
\subsubsection{In-waveguide Propagation}
Define $n_\mathrm{e}$ as the effective refractive index of the dielectric waveguide. The response vector from the feed point of the $n$th MW to its associated PAs is given by
\begin{equation} \label{gn}
	{\mathbf{g}_n}\left( {{\mathbf{x}_n}} \right) = {\left[ {{\rho_{n,1}e^{ - \mathrm{j}\frac{{2\pi }}{{{\lambda _\mathrm{W}}}}{x_{n,1}}}}, \ldots ,{\rho_{n,L}e^{ - \mathrm{j}\frac{{2\pi }}{{{\lambda _\mathrm{W}}}}{x_{n,L}}}}} \right]^T} \in \mathbb{C}^{L \times 1},
\end{equation}
where $\lambda _\mathrm{W}={\lambda}/{n_\mathrm{e}}$, with $\lambda$ denoting the free-space wavelength. $\rho_{n,l}$ denotes a power-scaling factor of the $l$th PA. For simplicity, we assume that the transmit power is equally radiated by PAs along the waveguide, which implies $\rho_{n,l}={1}/{\sqrt L}$, $\forall 1 \le n \le N$, $\forall 1 \le l \le L$ \cite{PASS3}.
\subsubsection{Free-space Propagation}
We consider high-frequency bands such as millimeter wave (mmWave). Thus, the LoS path dominates, and the non-LoS (NLoS) component can be neglected due to severe path loss and shadowing at the high-frequency bands. We adopt the general geometric free-space spherical wavefront model, which is applicable to both far- and near-field propagations. The channel vector from the PAs on the $n$th MW to the $k$th user can be expressed as
\begin{align}\label{los}
	&{\mathbf{h}}_{k,n}^H\left( \mathbf{x}_n, y_n\right) \nonumber\\
	&= \sqrt \beta  \left[ {\frac{{{e^{ - {\text{j}}\frac{{2\pi }}{\lambda }{{\left\| {{{\mathbf{u}}_{n,1}} - {\mathbf{u}}_k^{\text{U}}} \right\|}_2}}}}}{{{{\left\| {{{\mathbf{u}}_{n,1}} - {\mathbf{u}}_k^{\text{U}}} \right\|}_2}}}, \ldots ,\frac{{{e^{ - {\text{j}}\frac{{2\pi }}{\lambda }{{\left\| {{{\mathbf{u}}_{n,L}} - {\mathbf{u}}_k^{\text{U}}} \right\|}_2}}}}}{{{{\left\| {{{\mathbf{u}}_{n,L}} - {\mathbf{u}}_k^{\text{U}}} \right\|}_2}}}} \right] \in {\mathbb{C}^{1 \times L}},
\end{align}
where $\beta=\frac{\lambda^2}{16 \pi^2}$ represents the reference channel gain at 1 meter (m). 

By jointly considering in-waveguide propagation and free-space propagation, the channel vector from the BS to the $k$th user, denoted by $\mathbf{h}_k^H\left( \mathbf{X},\mathbf{y}\right) \in {\mathbb{C}^{1 \times N}}$, can be expressed as
\begin{align}\label{hk}
	& \mathbf{h}_k^H\left( \mathbf{X},\mathbf{y}\right)  \nonumber\\
	&= \left[ \mathbf{h}_{k,1}^H\left( \mathbf{x}_1, y_1\right)\mathbf{g}_1\left( \mathbf{x}_1\right), \ldots, \mathbf{h}_{k,N}^H\left( \mathbf{x}_N, y_N\right)\mathbf{g}_N\left( \mathbf{x}_N\right)\right]  .
\end{align}
\subsection{Signal Model}
The transmitted signals are multiplexed at the baseband via transmit beamforming/precoding, converted by the RF chains, and fed into the MWs for radiation. Let ${\mathbf{w}_k} \in {\mathbb{C}^{N \times 1}}$ denote the transmit beamforming vector for the $k$th user. The received signal at the $k$th user can be expressed as 
\begin{equation}
	{r_k} = {\mathbf{h}}_k^H\left( \mathbf{X},\mathbf{y}\right){\mathbf{w}_k}s_k + \sum\limits_{i = 1,i \ne k}^K {{\mathbf{h}}_k^H\left( \mathbf{X},\mathbf{y}\right){\mathbf{w}_i}s_i}  + {n_k},
\end{equation}
where $\mathbf{s}=\left[s_1,\ldots,s_K \right]^T\in \mathbb{C}^{K\times 1} $ denotes the normalized information-bearing signals satisfying $\mathbb{E}\left(\mathbf{s}\mathbf{s}^H \right)=\mathbf{I}_K $, and $n_k \sim \mathcal{CN}\left(0, \sigma ^2\right) $ represents the additive Gaussian noise at the $k$th user with average power $\sigma ^2$. Then, the achievable rate of the $k$th user is given by
\begin{equation}\label{rate}
	{R _k} =\log_2\left(1+  \frac{{{{\left| {\mathbf{h}_k^H\left( \mathbf{X},\mathbf{y}\right){\mathbf{w}_k}} \right|}^2}}}{{\sum\limits_{i = 1,i \ne k}^{{K}} {{{\left| {\mathbf{h}_k^H\left( \mathbf{X},\mathbf{y}\right){\mathbf{w}_i}} \right|}^2}}   + \sigma^2}}\right).
\end{equation}
\subsection{Problem Formulation}
To guarantee user fairness, we aim to maximize the minimum rate among all users by jointly optimizing the PA positions, the MW positions, and the beamforming vectors, which can be formulated as\footnote{This paper assumes perfect channel state information (CSI) to characterize the performance upper bound of the proposed MW-enabled PASS, where the CSI can be estimated from sparse channel measurements using, e.g., the orthogonal matching pursuit (OMP) algorithm \cite{channel}. Nevertheless, the impact of imperfect CSI on the considered system will be evaluated via simulations in Section \ref{5}.}
\begin{subequations}
	\label{max1}
	\begin{align}
		& \mathop {\mathrm{max} }\limits_{\mathbf{X},\mathbf{y},\left\{\mathbf{w}_k\right\}_{k=1}^K}  \mathop {\min }\limits_{1 \le k \le K} {R _k}\\
		&\mathrm{s.t.} \quad  \sum\limits_{k = 1}^{{K}} {\left\| {{\mathbf{w}_k}} \right\|_2^2}   \le P,\label{c1}\\
		&\hspace{2.3em}  0 \le x_{n,l} \le S_\mathrm{x}, \forall 1 \le n \le N,\forall 1 \le l \le L,\label{c2}\\
		&\hspace{2.3em} 0 \le y_n \le S_\mathrm{y},  \forall 1 \le n \le N,\label{c3}\\
		&\hspace{2.3em}  x_{n,l}-x_{n,l-1} \ge D_\mathrm{min}^\mathrm{PA},  \forall 1 \le n \le N,\forall 1<l\le L, \label{c4}\\
		&\hspace{2.3em} y_n-y_{n-1} \ge D_\mathrm{min}^\mathrm{MW},  \forall 1<n\le N, \label{c5}
	\end{align}
\end{subequations}	
where constraint \eqref{c1} indicates that the transmit power of the BS cannot exceed its power budget $P$. Constraints \eqref{c2} and \eqref{c3} respectively confine the PA and MW positions within finite ranges. Constraints \eqref{c4} and \eqref{c5} ensure the minimum inter-PA spacing $D_\mathrm{min}^\mathrm{PA}$ and the minimum inter-MW spacing $D_\mathrm{min}^\mathrm{MW}$, respectively, in order to avoid mutual coupling. Problem \eqref{max1} is challenging to solve because it is a non-convex optimization problem with coupled variables. Thus, we first address the special case of the single-user scenario, and then extend the results to the general multi-user case. 

\section{Single-User System}\label{3}
This section considers the single-user setup, i.e., $K=1$. For brevity, the user index $k$ is omitted. In this case, the maximum ratio transmission (MRT) is optimal, as there is no interference for this single user. Accordingly, the optimal transmit beamforming vector is given by
\begin{equation}
	\mathbf{w} = \sqrt P \frac{\mathbf{h}\left( \mathbf{X},\mathbf{y}\right)}{{{{\left\| \mathbf{h} \left( \mathbf{X},\mathbf{y}\right)\right\|}_2}}}.
\end{equation}
Thus, the achievable rate of the user is simplified to 
\begin{equation}
	R={\log _2}\left( {1 + \frac{{P\left\| \mathbf{h}\left( \mathbf{X},\mathbf{y}\right) \right\|_2^2}}{{{\sigma ^2}}}} \right).
\end{equation}
The achievable rate maximization problem can be equivalently reformulated as a channel gain maximization problem, i.e.,
\begin{equation}\label{max2}
 \mathop {\mathrm{max} }\limits_{\mathbf{X},\mathbf{y}} \quad \left\| \mathbf{h}\left( \mathbf{X},\mathbf{y}\right) \right\|_2^2\quad \mathrm{s.t.} \quad  \eqref{c2},\eqref{c3},\eqref{c4},\eqref{c5}. 
\end{equation}
In the following, we first consider the special case where each MW is equipped with a single PA and derive the optimal solutions for the MW and PA positions. Subsequently, for the general case with multiple PAs per MW, we develop an efficient algorithm to solve problem \eqref{max2}.
\subsection{Single Pinching Antenna per Movable Waveguide}\label{singlePA}
\begin{figure}[!t]
	\centering
	\includegraphics[width=1\linewidth]{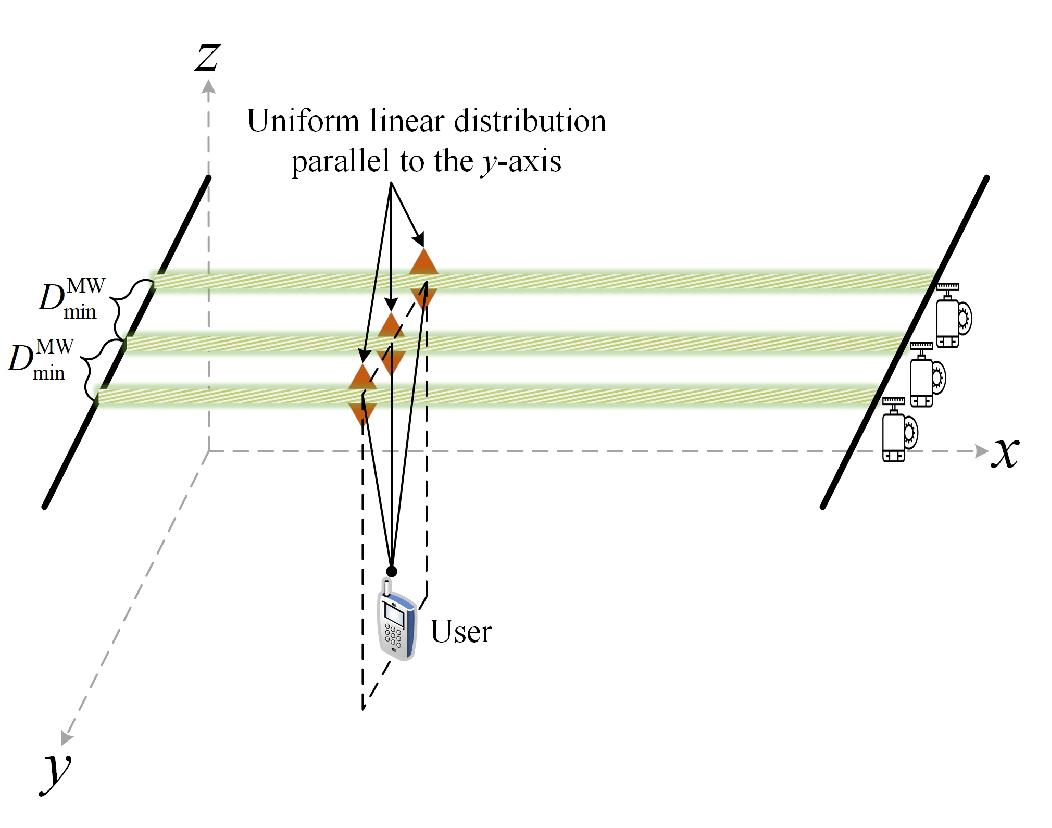}
	\caption{Illustration of the optimal MW and PA positions for the case of a single PA per MW.}
	\label{K1L1}
\end{figure}
When each MW contains only one PA, i.e., $L=1$, the channel gain in the objective function of problem \eqref{max2} can be expressed as
\begin{align}\label{Gsul1}
	G_{L = 1}^\mathrm{SU} & = {\sum\limits_{n = 1}^N {\left| {\sqrt \beta  \frac{{{e^{ - \mathrm{j}2\pi \left( {\frac{1}{\lambda }{{\left\| {{\mathbf{u}_n} - {\mathbf{u}^\mathrm{U}}} \right\|}_2} + \frac{1}{{{\lambda _\mathrm{W}}}}{x_n}} \right)}}}}{{{{\left\| {{\mathbf{u}_n} - {\mathbf{u}^\mathrm{U}}} \right\|}_2}}}} \right|} ^2} \nonumber\\
	&= {\sum\limits_{n = 1}^N {\left| {\sqrt \beta  \frac{1}{{{{\left\| {{\mathbf{u}_n} - {\mathbf{u}^\mathrm{U}}} \right\|}_2}}}} \right|} ^2},
\end{align}
where the PA index $l$ is omitted for brevity. We can see in \eqref{Gsul1} that the phase variations induced by different MW and PA positions do not alter the channel gain. To maximize the channel gain, it is sufficient to move the MWs and the PAs as close as possible to the user to minimize the large-scale path loss. As shown in Fig. \ref{K1L1}, the optimal $y$-axis coordinates of the MW positions can be obtained as
\begin{equation}\label{yn}
	{y_n} = {{y}^\mathrm{U}} - \frac{{N - 1}}{2}{D_{\min }^\mathrm{MW}} + \left( {n - 1} \right){D_{\min }^\mathrm{MW}}, \forall 1\le n\le N,
\end{equation}
which indicates that the MWs are arranged symmetrically, centered at the user's $y$-axis coordinate ${{y}^\mathrm{U}}$, with spacing $D_{\min }^\mathrm{MW}$. The optimal $x$-axis coordinates of the PA positions are given by 
\begin{equation}
	x_n = x^\mathrm{U},\forall 1\le n\le N,
\end{equation}
which implies that the PAs are located on a straight line directly above the user and parallel to the $y$-axis.
\subsection{Multiple Pinching Antennas per Movable Waveguide}\label{sec_su_mPA}
When multiple PAs are deployed on each MW, i.e., $L>1$, the channel gain is given by 
\begin{align}\label{gl2}
	G_{L > 1}^\mathrm{SU} &={\sum\limits_{n = 1}^N {\left| {\mathbf{h}_n^H\left( {{\mathbf{x}_n},{y_n}} \right){\mathbf{g}_n}\left( {{\mathbf{x}_n}} \right)} \right|} ^2}\nonumber\\
	&= {\sum\limits_{n = 1}^N {\left| {\sum\limits_{l = 1}^L {\sqrt {\frac{\beta }{L}} \frac{{{e^{ - \mathrm{j}2\pi \left( {\frac{1}{\lambda }{{\left\| {{\mathbf{u}_{n,l}} - {\mathbf{u}^\mathrm{U}}} \right\|}_2} + \frac{1}{{{\lambda _\mathrm{W}}}}{x_{n,l}}} \right)}}}}{{{{\left\| {{\mathbf{u}_{n,l}} - {\mathbf{u}^\mathrm{U}}} \right\|}_2}}}} } \right|} ^2}.
\end{align}
Thus, the positions of the MWs and PAs not only determine the small-scale phase variations but also affect the large-scale path loss. It is noted that if the phase of the free-space propagation response vector aligns with that of the in-waveguide propagation response vector, which is termed as the phase alignment condition, i.e., 
\begin{equation}\label{phase}
	{\mathbf{g}_n}\left( {{\mathbf{x}_n}} \right) = \frac{1}{{\sqrt L }}{e^{\mathrm{j}\angle {\mathbf{h}_n}\left( {{\mathbf{x}_n},{y_n}} \right)}},\forall 1\le n\le N,
\end{equation}
and the distance between the PAs and the user is minimized, which is termed as the minimum distance condition, i.e.,
\begin{equation}\label{min_dis}
	{\left\| {{\mathbf{u}_{n,l}} - {\mathbf{u}^\mathrm{U}}} \right\|_2} = h,\forall 1\le n\le N,\forall 1\le l\le L,
\end{equation}
the channel gain in \eqref{gl2} is upper-bounded by
\begin{align}\label{channel_ub}
	G_{L > 1}^\mathrm{SU} &= {\sum\limits_{n = 1}^N {\left| {\sum\limits_{l = 1}^L {\sqrt {\frac{\beta }{L}} \frac{{{e^{ - \mathrm{j}2\pi \left( {\frac{1}{\lambda }{{\left\| {{\mathbf{u}_{n,l}} - {\mathbf{u}^\mathrm{U}}} \right\|}_2} + \frac{1}{{{\lambda _\mathrm{W}}}}{x_{n,l}}} \right)}}}}{{{{\left\| {{\mathbf{u}_{n,l}} - {\mathbf{u}^\mathrm{U}}} \right\|}_2}}}} } \right|} ^2}\nonumber\\
	&\mathop  \le \limits^{\left( {{\mathrm{a}_1}} \right)} {\sum\limits_{n = 1}^N {\left( {\sum\limits_{l = 1}^L {\sqrt {\frac{\beta }{L}} \frac{1}{{{{\left\| {{\mathbf{u}_{n,l}} - {\mathbf{u}^\mathrm{U}}} \right\|}_2}}}} } \right)} ^2}\mathop  \le \limits^{\left( {{\mathrm{a}_2}} \right)} \frac{{NL\beta }}{{{h^2}}},
\end{align}
where the equality holds in ($\mathrm{a}_1$) if the phase alignment condition is satisfied, and the equality holds in ($\mathrm{a}_2$) if the minimum distance condition is satisfied. However, the simultaneous satisfaction of the phase alignment and minimum distance conditions is generally difficult to achieve in practice, since the user is randomly distributed and the minimum inter-PA spacing constraint \eqref{c4} prohibits placing all PAs at the same position directly above the user. Moreover, for the general case where multiple PAs are deployed on each MW, an exhaustive search to determine the optimal MW and PA positions leads to prohibitively high computational complexity. Therefore, we develop a two-stage algorithm with low complexity to efficiently solve problem \eqref{max2} as follows.

Since the MW positions $\mathbf{y}$ and the PA positions $\mathbf{X}$ are highly coupled in the objective function of problem \eqref{max2}, we decompose problem \eqref{max2} into two subproblems: 1) MW position optimization and 2) PA position optimization.
\subsubsection{MW Position Optimization}
In the first stage, for the given PA positions $\mathbf{X}$, we optimize the MW positions to minimize the large-scale path loss, which corresponds to satisfying the minimum distance condition in \eqref{min_dis} as much as possible without considering the phase alignment condition in \eqref{phase}. Accordingly, the MW position optimization problem is formulated as
\begin{equation}\label{max3}
	\mathop {\mathrm{min} }\limits_{\mathbf{y}} \quad  {\sum\limits_{n = 1}^N { { \left| {{y_n} - {y^\mathrm{U}}} \right| } } }\quad \mathrm{s.t.} \quad  \eqref{c3},\eqref{c5}.
\end{equation}
Based on the insights provided in Section \ref{singlePA}, the optimal solution to problem \eqref{max3} is given in \eqref{yn}.
\subsubsection{PA Position Optimization}
In the second stage, for the given MW positions $\mathbf{y}$, the PA position optimization problem can be formulated as
\begin{equation}\label{max4}
	\mathop {\mathrm{max} }\limits_{\mathbf{X}} \quad  \left\| \mathbf{h}\left( \mathbf{X},\mathbf{y}\right) \right\|_2^2 \quad \mathrm{s.t.} \quad  \eqref{c2},\eqref{c4}. 
\end{equation}
Intuitively, we can obtain the optimal PA positions that further minimize the large-scale path loss as $x_{n,l}=x^\mathrm{U}$, $\forall 1 \le n \le N$, $\forall 1 \le l \le L$. Based on this insight, we propose a sequential optimization algorithm with 1D local search to satisfy constraint \eqref{c4} and adjust the small-scale phases of the channel responses. First, with the user's $x$-axis coordinate $x^\mathrm{U}$ as the center, the interval $\left[ {{x^\mathrm{U}} - \frac{\Delta }{2},{x^\mathrm{U}} + \frac{\Delta }{2}} \right]$ is discretized into $Q$ points. Here, $\Delta$ denotes the interval length, which is typically at the wavelength scale to ensure that the variations of the PA positions within this range have a negligible effect on the large-scale path loss. Then, the set of candidate positions is defined as
\begin{equation}
	\mathcal{X}\left( \mathcal{Q}\right)   \triangleq \min\left(  \left\{ {\max \left( {{x^\mathrm{U}} - \frac{\Delta }{2},0} \right) + \frac{\Delta }{{Q - 1}}q} \right\},S_\mathrm{x}\right) ,
\end{equation}
where $q \in \mathcal{Q}$ is the position index and $\mathcal{Q} \triangleq \left\{ {0, \ldots ,Q - 1} \right\}$ denotes the set of position indices. Considering the minimum inter-PA spacing constraint \eqref{c4}, the set of feasible positions for the $l$th PA on the $n$th MW is given by
\begin{equation}
	{\mathcal{X} _{n,l}} = \mathcal{X}( \mathcal{Q}\backslash\tilde{\mathcal{Q}}_{n,l}) ,
\end{equation}
where $\tilde{\mathcal{Q}}_{n,l}$ is the set of infeasible position indices, defined as
\begin{equation}
	\tilde {\mathcal{Q}}_{n,l} \triangleq \left\{ {q\left| {\mathcal{X} \left( \mathcal{Q}\right) - {x_{n,\tilde l}} < {D_{\min }^\mathrm{PA}},\forall1 \le \tilde l \ne l \le L} \right.} \right\}.
\end{equation}
With all the other PA positions fixed, the channel gain in \eqref{gl2} can be reformulated with respect to the single variable $x_{n,l}$ as
\begin{align}
	&G_{L > 1}^\mathrm{SU}\left( {{x_{n,l}}} \right) \nonumber\\
	&= \frac{\beta }{L}{\left| {\frac{{{e^{ - \mathrm{j}2\pi \left( {\frac{1}{\lambda }\sqrt {{{\left( {{x_{n,l}} - {x^\mathrm{U}}} \right)}^2} + {c_1}}  + \frac{1}{{{\lambda _\mathrm{W}}}}{x_{n,l}}} \right)}}}}{{\sqrt {{{\left( {{x_{n,l}} - {x^\mathrm{U}}} \right)}^2} + {c_1}} }} + {c_2}} \right|^2} + {c_3},
\end{align}
where
\begin{subequations}
\begin{align}
	{c_1} &= {\left( {{y_n} - {y^\mathrm{U}}} \right)^2} + {h^2} \in \mathbb{R}, \\
	{c_2} &= \sum\limits_{\tilde l = 1,\tilde l \ne l}^L {\frac{{{e^{ - \mathrm{j}2\pi \left( {\frac{1}{\lambda }{{\left\| {{\mathbf{u}_{n,\tilde l}} - {\mathbf{u}^\mathrm{U}}} \right\|}_2} + \frac{1}{{{\lambda _\mathrm{W}}}}{x_{n,\tilde l}}} \right)}}}}{{{{\left\| {{\mathbf{u}_{n,\tilde l}} - {\mathbf{u}^\mathrm{U}}} \right\|}_2}}}} \in \mathbb{C}, \\
	{c_3} &= {\sum\limits_{\tilde n = 1,\tilde n \ne n}^N {\left| {\sum\limits_{l = 1}^L {\sqrt {\frac{\beta }{L}} \frac{{{e^{ - \mathrm{j}2\pi \left( {\frac{1}{\lambda }{{\left\| {{\mathbf{u}_{\tilde n,l}} - {\mathbf{u}^\mathrm{U}}} \right\|}_2} + \frac{1}{{{\lambda _\mathrm{W}}}}{x_{\tilde n,l}}} \right)}}}}{{{{\left\| {{\mathbf{u}_{\tilde n,l}} - {\mathbf{u}^\mathrm{U}}} \right\|}_2}}}} } \right|} ^2}\in \mathbb{R},
\end{align}
\end{subequations}
are three constants independent of $x_{n,l}$. Therefore, for each candidate PA position, it is not necessary to calculate the complete channel gain in the objective function of problem \eqref{max4}; instead, only the component associated with $x_{n,l}$ needs to be calculated, thereby further reducing the computational complexity. By local exhaustive search, the approximately optimal PA position $x_{n,l}$ can be obtained as 
\begin{equation}\label{xnl}
	{x_{n,l}} = \arg \mathop {\max }\limits_{{x_{n,l}} \in {\mathcal{X} _{n,l}}} \left| {\frac{{{e^{ - \mathrm{j}2\pi \left( {\frac{1}{\lambda }\sqrt {{{\left( {{x_{n,l}} - {x^\mathrm{U}}} \right)}^2} + {c_1}}  + \frac{1}{{{\lambda _\mathrm{W}}}}{x_{n,l}}} \right)}}}}{{\sqrt {{{\left( {{x_{n,l}} - {x^\mathrm{U}}} \right)}^2} + {c_1}} }} + {c_2}} \right|.
\end{equation}
\subsection{Overall Algorithm}
\begin{algorithm}[!t]
	\caption{Proposed algorithm for solving problem \eqref{max2}}
	\label{alg1}
	\renewcommand{\algorithmicrequire}{\textbf{Input:}}
	\renewcommand{\algorithmicensure}{\textbf{Output:}}
	\begin{algorithmic}[1]
		\REQUIRE $N$, $L$, $\lambda$, $h$, $S_\mathrm{x}$, $S_\mathrm{y}$, $n_\mathrm{e}$, $\sigma^2$, $P$, $D_{\min}^\mathrm{MW}$, $D_{\min}^\mathrm{PA}$, $Q$, $\Delta$, $\mathbf{u}^\mathrm{U}$, $\epsilon$.
		\ENSURE $\mathbf{X}$, $\mathbf{y}$.
		\STATE Update the optimal MW positions $\mathbf{y}$ by \eqref{yn}.
		\REPEAT
		\FOR{$n=1:1:N$}
		\FOR{$l=1:1:L$}
		\STATE Update $x_{n,l}$ by \eqref{xnl} through 1D local search.
		\ENDFOR
		\ENDFOR
		\UNTIL {Increase in the objective function of problem \eqref{max2} is less than $\epsilon$}
		\RETURN $\mathbf{X}$, $\mathbf{y}$.
	\end{algorithmic}
\end{algorithm}
The overall algorithm for solving problem \eqref{max2} is detailed in Algorithm \ref{alg1}. In line 1, the optimal MW positions are obtained by \eqref{yn}. In lines 2-8, the PA positions are alternately optimized based on \eqref{xnl} via the 1D local search, and the iteration is terminated if the increase in the objective function of problem \eqref{max2} is less than a predefined threshold $\epsilon$. Since the objective function of problem \eqref{max2} is non-decreasing over the iterations and is upper-bounded by a finite value, the convergence of Algorithm \ref{alg1} is guaranteed. As the MW positions are determined by a closed-form expression and the PA positions only need to be searched within a local interval of small length, the computational complexity of Algorithm \ref{alg1} is substantially lower than that of the exhaustive search algorithm or the alternating optimization algorithm over the entire waveguide length. Specifically, the complexity for calculating the optimal MW positions in line 1 is of order $\mathcal{O}\left(N \right) $. The complexity for iteratively updating the PA positions in lines 2-8 is of order $\mathcal{O}\left(INL Q\right) $, where $I$ denotes the number of iterations needed to achieve convergence. Thus, the total computational complexity of Algorithm \ref{alg1} is of order $\mathcal{O}\left(INLQ \right) $.
\section{Multi-User System}\label{4}
This section considers the general multi-user scenario, i.e., $K>1$. We first analyze the upper bounds on the objective function of problem \eqref{max1}. Then, we extend the algorithm for the single-user case to obtain an efficient solution to problem \eqref{max1}.
\subsection{Performance Analysis}
The achievable rate for each user in \eqref{rate} reaches an upper bound if the interference between any two users is nulled to zero while the received signal power is maximized. Specifically, the maximum received signal power for each user can be achieved by MRT, i.e., 
\begin{equation}
	{\mathbf{w}_k} = \sqrt {{p_k}} \frac{{{\mathbf{h}_k}\left( {\mathbf{X},\mathbf{y}} \right)}}{{{{\left\| {{\mathbf{h}_k}\left( {\mathbf{X},\mathbf{y}} \right)} \right\|}_2}}},
\end{equation}
where $p_k$ denotes the transmit power allocated to the $k$th user. Thus, the achievable rate in \eqref{rate} is upper-bounded by
\begin{align}\label{R1}
	{R_k}&\mathop  \le \limits^{\left( {{\mathrm{b}_1}} \right)} {\log _2}\left( {1 + \frac{{{{\left| {\mathbf{h}_k^H\left( {\mathbf{X},\mathbf{y}} \right){\mathbf{w}_k}} \right|}^2}}}{{{\sigma ^2}}}} \right)\nonumber\\
	&\mathop  \le \limits^{\left( {{\mathrm{b}_2}} \right)} {\log _2}\left( {1 + \frac{{{p_k}\left\| {{\mathbf{h}_k}\left( {\mathbf{X},\mathbf{y}} \right)} \right\|_2^2}}{{{\sigma ^2}}}} \right),
\end{align}
where the equality holds in ($\mathrm{b}_1$) if the channel vectors of any two users are orthogonal, i.e., $\mathbf{h}_k^H\left( {\mathbf{X},\mathbf{y}} \right){\mathbf{h}_{\tilde k}}\left( {\mathbf{X},\mathbf{y}} \right) = 0$, $\forall 1 \le k \ne \tilde k \le K$, and the equality holds in ($\mathrm{b}_2$) if the BS employs MRT to maximize the received signal power of the users. Furthermore, the channel gain of the $k$th user is upper-bounded by 
\begin{align}\label{channel_up}
	G_k^\mathrm{MU}& =\left\| {{\mathbf{h}_k}\left( {\mathbf{X},\mathbf{y}} \right)} \right\|_2^2={\sum\limits_{n = 1}^N {\left| {\mathbf{h}_{k,n}^H\left( {{\mathbf{x}_n},{y_n}} \right){\mathbf{g}_n}\left( {{\mathbf{x}_n}} \right)} \right|} ^2} \nonumber\\
	&\mathop  \le \limits^{\left( {{\mathrm{c}_1}} \right)} {\sum\limits_{n = 1}^N {\left( {\sum\limits_{l = 1}^L {\sqrt {\frac{\beta }{L}} \frac{1}{{{{\left\| {{\mathbf{u}_{n,l}} - \mathbf{u}_k^\mathrm{U}} \right\|}_2}}}} } \right)} ^2} \mathop  \le \limits^{\left( {{\mathrm{c}_2}} \right)}  \frac{{NL\beta }}{{{h^2}}},
\end{align}
which can be derived using a method similar to that in \eqref{channel_ub}, where the equality holds in ($\mathrm{c}_1$) if the phase alignment condition is satisfied, i.e., ${\mathbf{g}_n}\left( {{\mathbf{x}_n}} \right) = \frac{1}{{\sqrt L }}{e^{\mathrm{j}\angle {\mathbf{h}_{k,n}}\left( {{\mathbf{x}_n},{y_n}} \right)}}$, $\forall 1 \le k \le K$, $\forall 1 \le n \le N$, and the equality holds in ($\mathrm{c}_2$) if the minimum distance condition is satisfied, i.e., ${\left\| {{\mathbf{u}_{n,l}} - {\mathbf{u}_k^\mathrm{U}}} \right\|_2} = h$, $\forall 1 \le k \le K$, $\forall 1 \le n \le N$, $\forall 1 \le l \le L$. With the inequality ($\mathrm{c}_2$) in \eqref{channel_up}, an upper bound on the achievable rate of the $k$th user is obtained as 
\begin{equation}\label{R_ub}
	{R_k} \le {\log _2}\left( {1 + \frac{{{p_k}NL\beta }}{{{h^2}{\sigma ^2}}}} \right),
\end{equation}
where the power allocation should be $p_k = {P}/{K}$, $\forall 1 \le k \le K$, which maximizes the minimum upper bound across all $K$ users. Consequently, an upper bound on the minimum achievable rate can be derived in closed form as
\begin{equation}\label{Rup}
	{{R_\mathrm{ub}}}={{\log }_2}\left( {1 + \frac{{PNL\beta }}{{K{h^2}{\sigma ^2}}}} \right).
\end{equation} 

It is worth noting that the upper bound in \eqref{Rup} does not depend on the MW and PA positions and is unattainable in practice, because it is impossible for all PAs to simultaneously satisfy the minimum distance condition for all users. More practically, by ignoring the minimum distance condition and considering only the phase alignment condition, a tighter upper bound on the achievable rate of the $k$th user can be obtained using the inequality ($\mathrm{c}_1$) in \eqref{channel_up} as
\begin{align}\label{R_tub}
	{R_k} & \le {\log _2}\left( {1 + \frac{{{p_k}}}{{{\sigma ^2}}}{{\sum\limits_{n = 1}^N {\left( {\sum\limits_{l = 1}^L {\sqrt {\frac{\beta }{L}} \frac{1}{{{{\left\| {{\mathbf{u}_{n,l}} - \mathbf{u}_k^\mathrm{U}} \right\|}_2}}}} } \right)} }^2}} \right)\nonumber\\
	& = {\log _2}\left( {1 + \frac{{{p_k}}}{{L{\sigma ^2}}}\sum\limits_{n = 1}^N {\left\| {{\mathbf{h}_{k,n}}\left( {{\mathbf{x}_n},{y_n}} \right)} \right\|_1^2} } \right).
\end{align}
As shown in \eqref{R_tub}, the power allocation directly affects the upper bounds on the achievable rates of different users, and the equal transmit power allocation may not be optimal. To maximize the minimum achievable rate among all $K$ users, the transmit power should be allocated such that all users have the same achievable rate. Otherwise, the transmit power allocated to the users with higher achievable rates can always be reduced and reallocated to those with lower achievable rates to improve the max-min rate. Thus, the optimal power allocation for maximizing the minimum achievable rate can be obtained by solving the following equations:
\begin{equation}
	\left\{
	\begin{array}{l}
		{{{\log }_2}\left( {1 + \frac{{{p_k}}}{{L{\sigma ^2}}}
				\sum\limits_{n = 1}^N {\left\| {{\mathbf{h}_{k,n}}\left( {{\mathbf{x}_n},{y_n}} \right)} \right\|_1^2} } \right) 
			= R_\mathrm{ub}^\mathrm{tight}\left( {\mathbf{X},\mathbf{y}} \right)} ,\\
		{\sum\limits_{k = 1}^K {{p_k}}  = P},
	\end{array}
	\right.
\end{equation}
where $R_\mathrm{ub}^\mathrm{tight}\left( {\mathbf{X},\mathbf{y}} \right)$ denotes an upper bound on the minimum achievable rate, which is a function of the PA positions $\mathbf{X}$ and the MW positions $\mathbf{y}$, and can be derived as
\begin{align}\label{R_tup}
	& R_\mathrm{ub}^\mathrm{tight}\left( {\mathbf{X},\mathbf{y}} \right) \nonumber\\
	& = {\log _2}\left( {1 + \frac{P}{{L{\sigma ^2}}}{{\left( {\sum\limits_{k = 1}^K {{{\left( {\sum\limits_{n = 1}^N {\left\| {{\mathbf{h}_{k,n}}\left( {{\mathbf{x}_n},{y_n}} \right)} \right\|_1^2} } \right)}^{ - 1}}} } \right)}^{ - 1}}} \right).
\end{align}

Nevertheless, for certain channel conditions, the tight upper bound in \eqref{R_tup} cannot always be achieved. Hence, we develop a general algorithm with low complexity to obtain suboptimal solutions for problem \eqref{max1} in the following.
\subsection{Optimization Algorithm}\label{optimization}
With a sufficiently large antenna deployment region offered by PASS, the channel correlation among users can be effectively reduced via the joint optimization of MW and PA positions. As a result, we adopt the zero-forcing (ZF) method with low complexity for designing the transmit beamforming\footnote{Although the ZF beamforming is suboptimal, the simulations in Section \ref{5} will show that the ZF-based solution can achieve a performance close to the upper bound on the max-min rate. Moreover, the ZF beamforming has a closed-form expression, which reduces the computational complexity in MW and PA position optimizations. Therefore, the ZF-based solution offers a favorable trade-off between optimality and complexity.} to maximize the minimum achievable rate among all users. Define $\mathbf{H}\left( {\mathbf{X},\mathbf{y}} \right) \triangleq \left[ {{\mathbf{h}_1}\left( {\mathbf{X},\mathbf{y}} \right), \ldots ,{\mathbf{h}_K}\left( {\mathbf{X},\mathbf{y}} \right)} \right] \in {\mathbb{C}^{N \times K}}$. Considering the maximum transmit power constraint in \eqref{c1}, the power-normalized beamforming matrix is given by
\begin{align}\label{ZF}
	\mathbf{W} & = \left[ {{\mathbf{w}_1}, \ldots, {\mathbf{w}_K}} \right] \nonumber\\
	&= \sqrt P \frac{{\mathbf{H}\left( {\mathbf{X},\mathbf{y}} \right){{\left( {{\mathbf{H}^H}\left( {\mathbf{X},\mathbf{y}} \right)\mathbf{H}\left( {\mathbf{X},\mathbf{y}} \right)} \right)}^{ - 1}}}}{{{{\left\| {\mathbf{H}\left( {\mathbf{X},\mathbf{y}} \right){{\left( {{\mathbf{H}^H}\left( {\mathbf{X},\mathbf{y}} \right)\mathbf{H}\left( {\mathbf{X},\mathbf{y}} \right)} \right)}^{ - 1}}} \right\|}_F}}} \in \mathbb{C}^{N \times K},
\end{align}
which yields the same signal-to-noise ratio (SNR) for all $K$ users as 
\begin{equation}\label{sameSNR}
	\Gamma   = \frac{P}{{\mathrm{tr}\left( {{{\left( {{\mathbf{H}^H}\left( {\mathbf{X},\mathbf{y}} \right)\mathbf{H}\left( {\mathbf{X},\mathbf{y}} \right)} \right)}^{ - 1}}} \right){\sigma ^2}}}.
\end{equation}
Therefore, problem \eqref{max1} is reformulated as
\begin{align}\label{min1}
	& \mathop {\mathrm{min} }\limits_{\mathbf{X},\mathbf{y}} \quad \mathrm{tr}\left( {{{\left( {{\mathbf{H}^H}\left( {\mathbf{X},\mathbf{y}} \right)\mathbf{H}\left( {\mathbf{X},\mathbf{y}} \right)} \right)}^{ - 1}}} \right)\\
	&\mathrm{s.t.} \quad  \eqref{c2},\eqref{c3},\eqref{c4},\eqref{c5}. \nonumber
\end{align}
To address this non-convex optimization problem, we propose a two-scale search algorithm, in which the large-scale global search first determines the coarse MW and PA positions, while the small-scale local search then refines them around their optimized coarse positions.
\subsubsection{Large-Scale Global Search}\label{large}
At this stage, the linear moving regions of the MWs and the PAs along the $y$-axis and the $x$-axis are discretized into $Q_\mathrm{y}^\mathrm{L}$ and $Q_\mathrm{x}^\mathrm{L}$ points, respectively. Then, the sets of candidate MW and PA positions are respectively defined as
\begin{subequations}
\begin{align}
	\mathcal{Y}^\mathrm{L} & \triangleq \left\{ {\frac{{{S_\mathrm{y}}}}{{{Q_\mathrm{y}^\mathrm{L}} - 1}}{q_\mathrm{y}^\mathrm{L}}} \right\},{q_\mathrm{y}^\mathrm{L}} \in \left\{ {0, \ldots ,{Q_\mathrm{y}^\mathrm{L}}-1} \right\}, \\
	\mathcal{X}^\mathrm{L} & \triangleq \left\{ {\frac{{{S_\mathrm{x}}}}{{{Q_\mathrm{x}^\mathrm{L}} - 1}}{q_\mathrm{x}^\mathrm{L}}} \right\},  {q_\mathrm{x}^\mathrm{L}} \in \left\{ {0, \ldots ,{Q_\mathrm{x}^\mathrm{L}-1}} \right\}.
\end{align}
\end{subequations}
To alleviate the computational burden, these discretized positions are sparsely spaced, and we do not consider the minimum spacing constraints \eqref{c4} and \eqref{c5} at this stage.

First, the positions of all MWs are sequentially optimized based on the given PA positions. By fixing the remaining MWs, the position of the $n$th MW can be obtained as 
\begin{equation}\label{MW_pos}
	{y_n} = \arg \mathop {\min }\limits_{{y_n} \in {\mathcal{Y}^\mathrm{L}}} \mathrm{tr}\left( {{{\left( {{\mathbf{H}^H}\left( {\mathbf{X},\mathbf{y}} \right)\mathbf{H}\left( {\mathbf{X},\mathbf{y}} \right)} \right)}^{ - 1}}} \right).
\end{equation}
It is worth noting that calculating the matrix inverse at each candidate position incurs high computational complexity, and the adjustment of the $n$th MW position does not alter all elements in $\mathbf{H}\left( {\mathbf{X},\mathbf{y}} \right)$. Accordingly, we reformulate the objective function of problem \eqref{min1} as
\begin{align}
	&\mathrm{tr}\left( {{{\left( {{\mathbf{H}^H}\left( {\mathbf{X},\mathbf{y}} \right)\mathbf{H}\left( {\mathbf{X},\mathbf{y}} \right)} \right)}^{ - 1}}} \right)\nonumber\\
	\mathop  = \limits^{\left( {{\mathrm{d}_1}} \right)} &\mathrm{tr}\left( {{{\left( {\tilde {\mathbf{h}}_n{{\tilde {\mathbf{h}}}^H_n} + {{\tilde {\mathbf{H}}}_n}} \right)}^{ - 1}}} \right)\nonumber\\
	\mathop  = \limits^{\left( {{\mathrm{d}_2}} \right)} &\mathrm{tr}\left( {{{ {{{\tilde {\mathbf{H}}_n}}} }^{ - 1}} - \frac{{{{ {{{\tilde {\mathbf{H}}}_n}} }^{ - 1}}\tilde {\mathbf{h}}_n{{\tilde {\mathbf{h}}}^H_n}{{ {{{\tilde {\mathbf{H}}}_n}} }^{ - 1}}}}{{1 + {{\tilde {\mathbf{h}}}^H_n}{{ {{{\tilde {\mathbf{H}}}_n}} }^{ - 1}}\tilde {\mathbf{h}}_n}}} \right),
\end{align}
where $\tilde {\mathbf{h}}_n \in \mathbb{C}^{K \times 1}$ denotes the $n$th column of ${\mathbf{H}^H}\left( {\mathbf{X},\mathbf{y}} \right)$. The equality ($\mathrm{d}_1$) holds by defining ${{\tilde {\mathbf{H}}}_n} = \sum\nolimits_{\tilde n = 1,\tilde n \ne n}^N {\tilde {\mathbf{h}}_{\tilde n}{{\tilde {\mathbf{h}}}^H_{\tilde n}}} \in \mathbb{C}^{K \times K}$, and the equality ($\mathrm{d}_2$) holds by applying the Sherman-Morrison lemma \cite{PASS9}. Thus, \eqref{MW_pos} can be reformulated as
\begin{equation}\label{MW_pos1}
	{y_n} = \arg \mathop {\max }\limits_{{y_n} \in {\mathcal{Y}^\mathrm{L}}} \frac{{{{\tilde {\mathbf{h}}}^H_n\left(y_n \right) }\tilde {\mathbf{H}}_n^{ - 2}\tilde {\mathbf{h}}_n\left(y_n \right)}}{{1 + {{\tilde {\mathbf{h}}}^H_n\left(y_n \right)}\tilde {\mathbf{H}}_n^{ - 1}\tilde {\mathbf{h}}_n\left(y_n \right)}}.
\end{equation}
By sequentially updating $\left\{ {{y_n}} \right\}_{n = 1}^N$, the coarse MW positions can be obtained.

Then, the PA positions are optimized by adopting a similar sequential updating method based on the fixed MW positions. Specifically, the position of the $l$th PA on the $n$th MW can be selected as
\begin{equation}\label{PA_pos}
	{x_{n,l}} = \arg \mathop {\max }\limits_{{x_{n,l}} \in {\mathcal{X}^\mathrm{L}}} \frac{{{{\tilde {\mathbf{h}}}^H_n\left(x_{n,l} \right)}\tilde {\mathbf{H}}_n^{ - 2}\tilde {\mathbf{h}}_n\left(x_{n,l} \right)}}{{1 + {{\tilde {\mathbf{h}}}^H_n\left(x_{n,l} \right)}\tilde {\mathbf{H}}_n^{ - 1}\tilde {\mathbf{h}}_n\left(x_{n,l} \right)}}.
\end{equation}
With \eqref{PA_pos}, we can obtain the coarse positions of all PAs.
\subsubsection{Small-Scale Local Search}
To capture small-scale channel variations and satisfy the minimum spacing constraints \eqref{c4} and \eqref{c5}, this stage performs a small-scale local search for finely tuning the MW and PA positions based on their coarse positions obtained from \eqref{MW_pos1} and \eqref{PA_pos}, respectively. To determine the position of the $n$th waveguide, a small interval of length $\Delta_\mathrm{y}$, centered at $y_n$ in \eqref{MW_pos1}, is first densely discretized into $Q_\mathrm{y}^\mathrm{S}$ points, i.e.,
\begin{align}
	&\mathcal{Y}_{n}^\mathrm{S}\left( \mathcal{Q}_\mathrm{y}^\mathrm{S}\right)   \nonumber\\
	&\triangleq \min\left(  \left\{ {\max \left( {{y_{n}} - \frac{\Delta_\mathrm{y} }{2},0} \right) + \frac{\Delta_\mathrm{y} }{{Q_\mathrm{y}^\mathrm{S} - 1}}q_\mathrm{y}^\mathrm{S}} \right\},S_\mathrm{y}\right) ,
\end{align}
where $q_\mathrm{y}^\mathrm{S} \in \mathcal{Q}_\mathrm{y}^\mathrm{S} \triangleq \left\{0,\ldots,Q_\mathrm{y}^\mathrm{S} - 1\right\} $ denotes the position index. Due to the minimum inter-MW spacing constraint \eqref{c5}, we can define the set of infeasible position indices as
\begin{equation}
	\tilde {\mathcal{Q}}_{n}^\mathrm{S} \triangleq \left\{ {q\left| {\mathcal{Y}_{n}^\mathrm{S}\left( \mathcal{Q}_\mathrm{y}^\mathrm{S}\right) - {y_{\tilde n}} < {D_{\min }^\mathrm{MW}},\forall1 \le \tilde n \ne n \le N} \right.} \right\}.
\end{equation}
Thus, the set of feasible positions for the $n$th MW is given by
\begin{equation}
	{\tilde{\mathcal{Y}} _{n}^\mathrm{S}} = {\mathcal{Y}^\mathrm{S} _{n}}( \mathcal{Q}_\mathrm{y}^\mathrm{S}\backslash\tilde{\mathcal{Q}}_{n}^\mathrm{S}) .
\end{equation}
Then, the small-scale local search is carried out to refine the position $y_n$, i.e., 
\begin{equation}\label{refine_MW}
	{y_{n}} = \arg \mathop {\max }\limits_{{y_{n}} \in {\tilde{\mathcal{Y}} _{n}^\mathrm{S}}} \frac{{{{\tilde {\mathbf{h}}}^H_n\left(y_{n} \right)}\tilde {\mathbf{H}}_n^{ - 2}\tilde {\mathbf{h}}_n\left(y_{n} \right)}}{{1 + {{\tilde {\mathbf{h}}}^H_n\left(y_{n} \right)}\tilde {\mathbf{H}}_n^{ - 1}\tilde {\mathbf{h}}_n\left(y_{n} \right)}}.
\end{equation}
After finely updating each MW position with the others fixed, the final positions of all MWs are determined.

Similarly, the PA positions are refined based on the coarse positions obtained from \eqref{PA_pos}. Specifically, we discretize a small interval of length $\Delta_\mathrm{x}$ centered at $x_{n,l}$ in \eqref{PA_pos} into $Q_\mathrm{x}^\mathrm{S}$ points, i.e.,
\begin{align}
	&\mathcal{X}_{n,l}^\mathrm{S}\left( \mathcal{Q}_\mathrm{x}^\mathrm{S}\right)   \nonumber\\
	&\triangleq \min\left( \left\{ {\max \left( {{x_{n,l}} - \frac{\Delta_\mathrm{x} }{2},0} \right) + \frac{\Delta_\mathrm{x} }{{Q_\mathrm{x}^\mathrm{S} - 1}}q_\mathrm{x}^\mathrm{S}} \right\},S_\mathrm{x}\right) ,
\end{align}
where $q_\mathrm{x}^\mathrm{S} \in \mathcal{Q}_\mathrm{x}^\mathrm{S} \triangleq \left\{0,\ldots,Q_\mathrm{x}^\mathrm{S} - 1\right\} $ is the position index. Then, the set of feasible positions for the $l$th PA on the $n$th MW is given by
\begin{equation}
	{\tilde{\mathcal{X}} _{n,l}^\mathrm{S}} = {\mathcal{X}^\mathrm{S} _{n,l}}( \mathcal{Q}_\mathrm{x}^\mathrm{S}\backslash\tilde{\mathcal{Q}}_{n,l}^\mathrm{S}) ,
\end{equation}
where $\tilde{\mathcal{Q}}_{n,l}^\mathrm{S}$ is the set of infeasible position indices, defined as
\begin{equation}
	\tilde {\mathcal{Q}}_{n,l}^\mathrm{S} \triangleq \left\{ {q\left| {\mathcal{X}_{n,l}^\mathrm{S}\left( \mathcal{Q}_\mathrm{x}^\mathrm{S}\right) - {x_{n,\tilde l}} < {D_{\min }^\mathrm{PA}},\forall1 \le \tilde l \ne l \le L} \right.} \right\}.
\end{equation}
The exact PA position $x_{n,l}$ can be obtained through a local 1D search as
\begin{equation}\label{refine_PA}
	{x_{n,l}} = \arg \mathop {\max }\limits_{{x_{n,l}} \in {\tilde{\mathcal{X}} _{n,l}^\mathrm{S}}} \frac{{{{\tilde {\mathbf{h}}}^H_n\left(x_{n,l} \right)}\tilde {\mathbf{H}}_n^{ - 2}\tilde {\mathbf{h}}_n\left(x_{n,l} \right)}}{{1 + {{\tilde {\mathbf{h}}}^H_n\left(x_{n,l} \right)}\tilde {\mathbf{H}}_n^{ - 1}\tilde {\mathbf{h}}_n\left(x_{n,l} \right)}}.
\end{equation}

\begin{algorithm}[!t]
	\caption{Proposed algorithm for solving problem \eqref{max1}}
	\label{alg2}
	\renewcommand{\algorithmicrequire}{\textbf{Input:}}
	\renewcommand{\algorithmicensure}{\textbf{Output:}}
	\begin{algorithmic}[1]
		\REQUIRE $N$, $L$, $K$, $\lambda$, $h$, $S_\mathrm{x}$, $S_\mathrm{y}$, $n_\mathrm{e}$, $\sigma^2$, $P$, $D_{\min}^\mathrm{MW}$, $D_{\min}^\mathrm{PA}$, $Q_\mathrm{x}^\mathrm{L}$, $Q_\mathrm{x}^\mathrm{S}$, $Q_\mathrm{y}^\mathrm{L}$, $Q_\mathrm{y}^\mathrm{S}$, $\Delta_\mathrm{x}$, $\Delta_\mathrm{y}$, $\left\{\mathbf{u}^\mathrm{U}_k\right\}_{k=1}^K$, $\epsilon^\mathrm{L}$, $\epsilon^\mathrm{S}$.
		\ENSURE $\mathbf{X}$, $\mathbf{y}$.
		\REPEAT 
		\FOR{$n=1:1:N$}
		\STATE Coarsely update $y_n$ by \eqref{MW_pos1} through 1D global search. 
		\ENDFOR
		\FOR{$n=1:1:N$}
		\FOR{$l=1:1:L$}
		\STATE Coarsely update $x_{n,l}$ by \eqref{PA_pos} through 1D global search.
		\ENDFOR
		\ENDFOR
		\UNTIL{Increase in the objective function of problem \eqref{max1} is less than $\epsilon^\mathrm{L}$}
		\REPEAT 
		\FOR{$n=1:1:N$}
		\STATE Refine $y_n$ by \eqref{refine_MW} through 1D local search. 
		\ENDFOR
		\FOR{$n=1:1:N$}
		\FOR{$l=1:1:L$}
		\STATE Refine $x_{n,l}$ by \eqref{refine_PA} through 1D local search.
		\ENDFOR
		\ENDFOR
		\UNTIL{Increase in the objective function of problem \eqref{max1} is less than $\epsilon^\mathrm{S}$}
		\RETURN $\mathbf{X}$, $\mathbf{y}$.
	\end{algorithmic}
\end{algorithm}
The overall algorithm for solving problem \eqref{max1} is detailed in Algorithm \ref{alg2}. In lines 1-10, the coarse MW and PA positions are alternately updated until the increase in the objective function of problem \eqref{max1} falls below a predefined threshold $\epsilon^\mathrm{L}$, with a computational complexity of order $\mathcal{O}(I^\mathrm{L}NK^2(K+LQ_\mathrm{x}^\mathrm{L}+Q_\mathrm{y}^\mathrm{L}))$, where $I^\mathrm{L}$ denotes the number of iterations required for the large-scale search to achieve convergence. In lines 11-20, the exact MW and PA positions are iteratively determined until the increase in the objective function of problem \eqref{max1} is less than a predefined threshold $\epsilon^\mathrm{S}$, with a computational complexity of order $\mathcal{O}(I^\mathrm{S}NK^2(LQ_\mathrm{x}^\mathrm{S}+Q_\mathrm{y}^\mathrm{S}))$, where $I^\mathrm{S}$ denotes the number of iterations required for the small-scale search to achieve convergence. Thus, the total computational complexity of Algorithm \ref{alg2} is of order $\mathcal{O}(NK^2(I^\mathrm{L}(K+LQ_\mathrm{x}^\mathrm{L}+Q_\mathrm{y}^\mathrm{L})+I^\mathrm{S}(LQ_\mathrm{x}^\mathrm{S}+Q_\mathrm{y}^\mathrm{S})))$.
\subsection{Two-timescale Optimization}\label{two}
\begin{figure}[!t]
	\centering
	\includegraphics[width=1\linewidth]{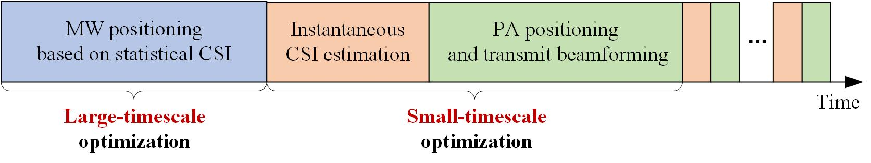}
	\caption{Illustration of the proposed two-timescale optimization scheme.}
	\label{block}
\end{figure}

In practice, the positioning of the MWs and PAs can be implemented in different ways. For MW positioning, each MW can be mounted on a mechanical slide and driven by an attached motor to adjust its position. Due to the extra mechanical positioning modules, MW repositioning may incur non-negligible response time and movement-induced energy consumption \cite{EE1}, which makes frequent MW position updates undesirable. By contrast, the PA positions can be updated by pre-installing multiple radiating points along each waveguide and electronically selecting a subset to activate, thereby enabling rapid position adaptation with much lower energy overhead. The joint optimization of the MW and PA positions in problem \eqref{max1} requires updating the MW positions based on instantaneous CSI, which implies that both channel estimation and MW movement need to be implemented within a time period shorter than the channel coherence time. Thus, this instantaneous CSI-based design mainly applies to systems with slowly varying channels. For fast-fading channels with rapidly moving users in a multi-path environment, the latency and energy cost incurred by MW movement may hinder the efficiency of the PASS. To address this issue, a two-timescale optimization scheme is proposed for the considered MW-enabled PASS, where the MW positions are designed based on statistical CSI on a large timescale, while the PA positions on the MWs and the beamforming vectors are optimized based on instantaneous CSI on a small timescale, as illustrated in Fig. \ref{block}. To account for sparse multi-path propagation in mmWave communication systems, we extend the free-space channel model with only a single LoS path in \eqref{los} to include both LoS and NLoS paths based on the Saleh-Valenzuela (SV) channel model \cite{SV,SV1}. Specifically, for the $k$th user, we consider $L_k$ NLoS paths, corresponding to $L_k$ effective point scatterers around the user, whose locations are denoted by $\mathbf{u}^{\mathrm{S}}_{k,\ell}=[x^{\mathrm{S}}_{k,\ell},y^{\mathrm{S}}_{k,\ell},z^{\mathrm{S}}_{k,\ell}]^{\mathrm{T}}$, $1\leq \ell \leq L_k$. The LoS distance between the $l$th PA on the $n$th MW and the $k$th user is given by $d_{k,n,l}^{\rm LoS}=||\mathbf{u}_{n,l}-\mathbf{u}_k^{\rm U}||_2$. The propagation distance of the $\ell$th NLoS path from the $l$th PA on the $n$th MW to the $k$th user via the corresponding scatterer is given by $d_{k,n,l,\ell}^{\rm NLoS}=||\mathbf{u}_{n,l}-\mathbf{u}_{k,\ell}^{\rm S}||_2+||\mathbf{u}_{k,\ell}^{\rm S}-\mathbf{u}_k^{\rm U}||_2$. Accordingly, the sparse NLoS component between the $l$th PA on the $n$th MW and the $k$th user is modeled as
\begin{equation}
	h_{k,n,l}^{\rm NLoS}
	=
	\sqrt{\frac{\beta}{L_k}}
	\sum_{\ell=1}^{L_k}
	\alpha_{k,\ell}
	\frac{
		e^{-\mathrm{j}\frac{2\pi}{\lambda}d_{k,n,l,\ell}^{\rm NLoS}}
	}{
		d_{k,n,l,\ell}^{\rm NLoS}
	},
	\label{eq:sparse_nlos_channel}
\end{equation}
where $\alpha_{k,\ell}\sim\mathcal{CN}\left( 0,1\right) $ denotes the complex gain of the $\ell$th effective scattering path. Then, the channel coefficient from the $l$th PA on the $n$th MW to the $k$th user can be expressed as
\begin{equation}
	h_{k,n,l}
	=
	\sqrt{\frac{\kappa_k}{\kappa_k+1}}
	\frac{\sqrt{\beta}
		e^{-\mathrm{j}\frac{2\pi}{\lambda}d_{k,n,l}^{\rm LoS}}
	}{
		d_{k,n,l}^{\rm LoS}
	}
	+
	\sqrt{\frac{1}{\kappa_k+1}}
	h_{k,n,l}^{\rm NLoS}
	,
	\label{eq:sv_scatterer_channel}
\end{equation}
where $\kappa_k$ denotes the LoS-to-NLoS power ratio. Therefore, the two-timescale optimization problem is formulated by maximizing the ergodic achievable rate of the $K$ users, which is given by
\begin{subequations}
	\label{max_exp}
	\begin{align}
		& \mathop {\mathrm{max} }\limits_{\mathbf{y}}
		\quad \mathbb{E}\left( \mathop {\mathrm{max} }\limits_{\mathbf{X},\left\{\mathbf{w}_k\right\}_{k=1}^K}  \mathop {\min }\limits_{1 \le k \le K} {R _k}\right)\label{obj_exp}\\
		&\mathrm{s.t.} \quad  \eqref{c1},\eqref{c2},\eqref{c3},\eqref{c4},\eqref{c5},\nonumber
	\end{align}
\end{subequations}
where the expectation is taken over all possible NLoS channel realizations. It is worth noting that problem \eqref{max_exp} represents a general formulation that is applicable to the MW-enabled PASS under arbitrary statistical CSI\footnote{In this paper, statistical CSI is determined by the spatial distribution of users, i.e., the user locations $\{\mathbf{u}^\mathrm{U}_k\}_{k=1}^K$, and is used to optimize the MW positions on the large timescale in the proposed two-timescale optimization scheme.
}. The MWs can be deployed at the designated positions for a long time period that is significantly longer than the coherence time of the instantaneous channels, thereby achieving a high ergodic achievable rate and reducing the energy consumption associated with MW movement. Within each transmission block, the instantaneous CSI acquisition, PA position optimization, and beamforming design developed for conventional PASS can be implemented with the optimized MW positions, and thus no additional overhead is introduced. In practical implementations, the MW movement frequency can be flexibly configured to balance the trade-off between energy consumption and communication performance.

To handle the intractable expectation in \eqref{obj_exp}, it is approximated via the Monte-Carlo method as
\begin{equation}\label{MC}
	\mathbb{E}\left( {\mathop {{\mathrm{max}}}\limits_{{\mathbf{X}},\left\{ {{{\mathbf{w}}_k}} \right\}_{k = 1}^K} \mathop {\min }\limits_{1 \le k \le K} {R_k}} \right) \approx \frac{1}{T}\sum\limits_{t = 1}^T {\mathop {{\mathrm{max}}}\limits_{{\mathbf{X}},\left\{ {{{\mathbf{w}}_k}} \right\}_{k = 1}^K} \mathop {\min }\limits_{1 \le k \le K} R_k^t} ,
\end{equation}
where $T$ is the total number of Monte-Carlo simulations, and $R_k^t$ denotes the achievable rate of the $k$th user for the $t$th independent channel realization. With a sufficiently large $T$, the channel statistics between the PASS and the users can be effectively captured. To solve problem \eqref{max_exp}, the ZF beamforming design in \eqref{ZF} can be applied. The MW positions are then optimized via the two-scale search algorithm in Section \ref{optimization}, where the objective is replaced by \eqref{MC}. After obtaining the long-term MW positions, the PA positions are further optimized in each transmission block using the same algorithm.
\section{Simulation Results} \label{5}
In this section, we provide the simulation results for the considered system. The simulation setup and benchmark schemes are first illustrated, and then the simulation results under single-user and multi-user scenarios are presented.
\subsection{Simulation Setup and Benchmark Schemes}
\begin{table}[!t]
	\renewcommand{\arraystretch}{1.5}
	\caption{Simulation Parameters}
	\label{tab1}
	\centering
	\begin{tabular}{|l|l|l|}
		\hline
		\textbf{Parameter} & \textbf{Description} & \textbf{Value} \\
		\hline
		$N$ & Number of waveguides & 16 \\
		\hline
		$L$ & Number of PAs on each waveguide & 4 \\
		\hline
		$K$ & Number of users & 6 \\
		\hline
		$L_k$ & Number of NLoS paths per user & 4 \\
		\hline
		$S_\mathrm{x}$, $ S_\mathrm{y}$ & Side lengths of the rectangular region & 100 m \\
		\hline
		$h$ & Height of the waveguides and PAs & 10 m \\
		\hline
		$\lambda$ & Carrier wavelength & 0.01 m \\
		\hline
		$n_\mathrm{e}$ & Effective refractive index & 1.4 \\
		\hline
		$P$ & Maximum transmit power & 30 dBm \\
		\hline
		$D_\mathrm{min}^\mathrm{MW}$ & Minimum inter-MW spacing & 0.5 m \\
		\hline
		$D_\mathrm{min}^\mathrm{PA}$ & Minimum inter-PA spacing & $\lambda/2$ \\
		\hline
		$\sigma^2$ & Average noise power & $-90$ dBm \\
		\hline
		$Q$ & 1D search resolution in Algorithm \ref{alg1} & $10^{3}$ \\
		\hline
		\makecell[l]{$Q_\mathrm{x}^\mathrm{L}$, $Q_\mathrm{x}^\mathrm{S}$,\\ $Q_\mathrm{y}^\mathrm{L}$, $Q_\mathrm{y}^\mathrm{S}$} & 
		1D search resolutions in Algorithm \ref{alg2} & $10^{2}$ \\
		\hline
		$\Delta$, $\Delta_\mathrm{x}$ & Interval lengths in Algorithms \ref{alg1} and \ref{alg2} & $10\lambda$ \\
		\hline
		$\Delta_\mathrm{y}$ & Interval length in Algorithm \ref{alg2} & 10 m \\
		\hline
		$\epsilon$, $\epsilon^\mathrm{L}$, $\epsilon^\mathrm{S}$ & Predefined thresholds in Algorithms \ref{alg1} and \ref{alg2} & $10^{-5}$ \\
		\hline
	\end{tabular}
\end{table}
In the simulations, the waveguides and PAs are positioned at the same height $h$. Without loss of generality, the users are assumed to be distributed within a rectangular region spanning $S_\mathrm{x} \times S_\mathrm{y}$. Unless otherwise specified, the default simulation parameters are listed in Table \ref{tab1}.

The results obtained by the proposed algorithms are termed as \textit{Proposed}. Besides, the following benchmark schemes are considered for performance comparison:
\begin{itemize}
	\item \textit{Upper bound}: The upper bound is given by ($\mathrm{a}_2$) in \eqref{channel_ub} for the single-user scenario and by \eqref{R_tup} for the multi-user scenario.
	
	\item \textit{Dense FPW}: The FPWs are densely deployed at the center of the rectangular region with a spacing of $D_{\min }^\mathrm{MW}$, i.e., 
	\begin{equation}\label{den}
		{y_n} = \frac{S_\mathrm{y}}{2} - \frac{{N - 1}}{2}{D_{\min }^\mathrm{MW}} + \left( {n - 1} \right){D_{\min }^\mathrm{MW}}, \forall 1\le n\le N.
	\end{equation}
	The PA positions are optimized using the scheme in Algorithm \ref{alg1} for the single-user case and the scheme in Algorithm \ref{alg2} for the multi-user case.
	\item \textit{Sparse FPW}: The FPWs are sparsely deployed over the range $\left[ 0, S_\mathrm{y}\right] $ along the $y$-axis, i.e., 
	\begin{equation}\label{spa}
		{y_n} =  {\frac{S_\mathrm{y}}{N-1}}\left( {n - 1} \right), \forall 1\le n\le N.
	\end{equation}
	The PA positions are also optimized using the scheme in Algorithm \ref{alg1} for the single-user case and the scheme in Algorithm \ref{alg2} for the multi-user case.
	\item \textit{Centralized FPA}: A uniform planar array (UPA) consisting of $N \times L$ FPAs with half-wavelength spacing is placed at the center of the rectangular region, i.e., $\left({S_\mathrm{x} }/{2},{S_\mathrm{y} }/{2}\right) $, with the same height $h$. Each FPA is connected to a dedicated RF chain, and thus this scheme requires more RF chains than the previous schemes.
	\item \textit{Distributed FPA}: A total of $N \times L$ FPAs are randomly distributed over the rectangular region at the same height $h$. Each FPA is connected to a dedicated RF chain, and thus this scheme also requires $N \times L$ RF chains.
\end{itemize}
\subsection{Single-User System}
\begin{figure}[!t]
	\centering
	\includegraphics[width=1\linewidth]{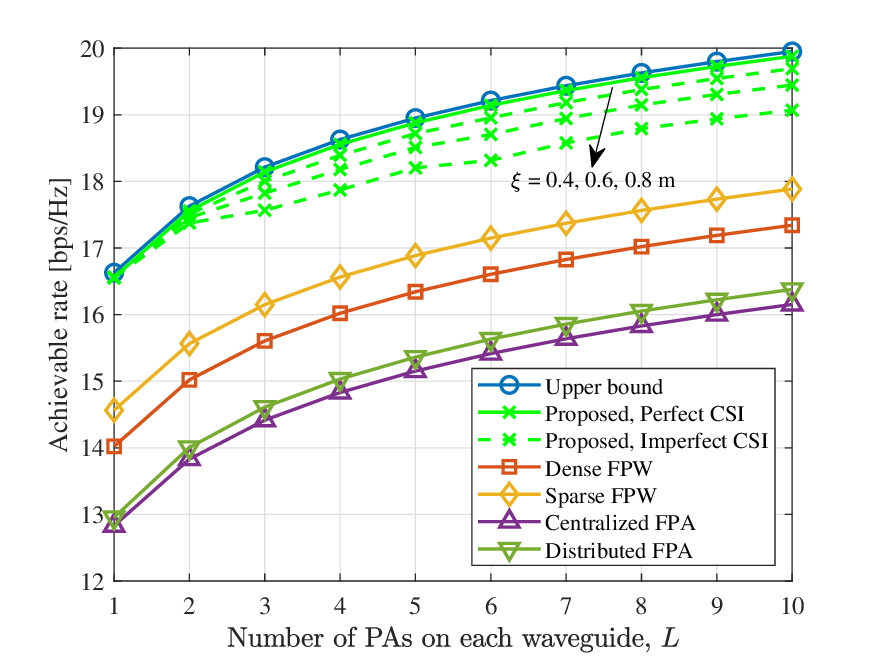}
	\caption{Achievable rate versus number of PAs on each waveguide.}
	\label{sg_L_rate}
\end{figure}

First, we consider the single-user setup, i.e., $K = 1$, where the user is randomly distributed within the rectangular region. Fig. \ref{sg_L_rate} shows the achievable rates versus the number of PAs $L$ on each waveguide. We can see that the performance of all schemes increases with $L$ because a larger $L$ yields a higher channel gain (see ($\mathrm{a}_2$) in \eqref{channel_ub}). Moreover, the PASS-based schemes outperform the two FPA schemes that employ more RF chains. This is because the fixed antenna deployment results in larger distances between the antennas and the user, whereas the PASS leverages the advantage of flexible antenna positioning to significantly reduce large-scale path loss, thereby improving the achievable rate for the user. Therefore, PASS offers the potential to reduce the costly RF chain overhead in practical applications while achieving better rate performance. In addition, the proposed scheme nearly achieves its upper bound and substantially outperforms the two FPW schemes, since the flexible movement of MWs enables PAs to better track the user at different locations. To evaluate the robustness of the proposed scheme to CSI errors induced by potential user mobility, Fig. \ref{sg_L_rate} further illustrates the system performance when the user randomly moves within a local circular region of radius $\xi$ m. As $\xi$ increases, the performance of the proposed scheme degrades only slightly, yet it still outperforms the other schemes even with perfect CSI.

\begin{figure}[!t]
	\centering
	\includegraphics[width=1\linewidth]{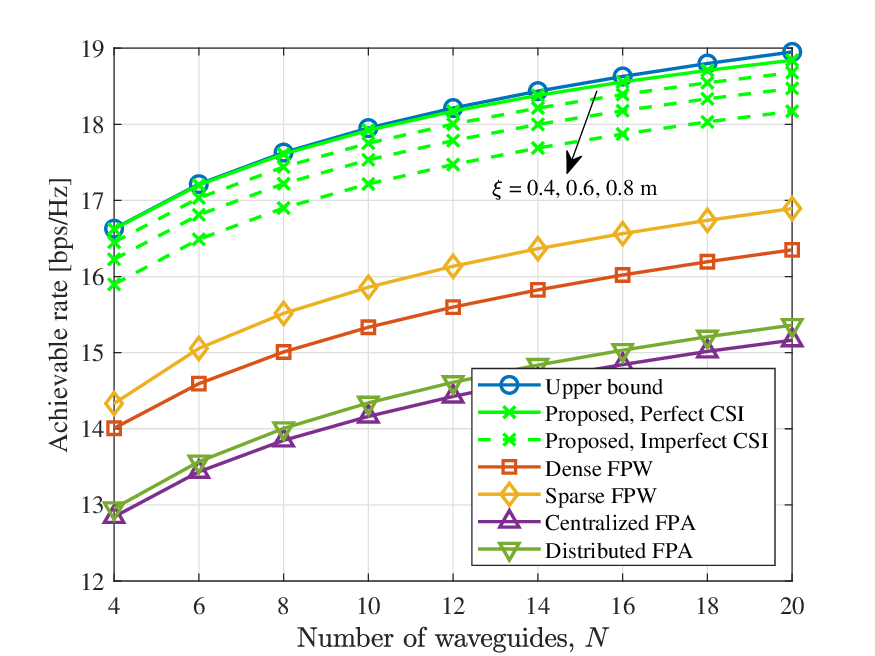}
	\caption{Achievable rate versus number of waveguides.}
	\label{sg_N_rate}
\end{figure}

Fig. \ref{sg_N_rate} investigates the impact of the number of waveguides $N$ on the achievable rates under various schemes. As $N$ increases, the achievable rate of the user improves for all schemes, since a larger number of RF chains enhances the beamforming gain. Furthermore, the sparse FPW scheme consistently achieves higher performance than the dense FPW scheme. This is because the minimum distance between the PAs and the user in the sparse FPW scheme can be significantly smaller due to the sparse deployment of waveguides. In other words, the sparse FPW scheme can take advantage of the PAs that are closer to the user and apply MRT to maximize the achievable rate. In contrast, in the dense FPW scheme, the PAs are nearly equidistant from the user and all located farther away, thus resulting in inferior performance. When $N$ increases from 4 to 20, the performance gap between the two FPW schemes and that between the two FPA schemes increase by 67\% and 78\%, respectively, which indicates that the distributed antenna deployment is more suitable than the centralized deployment for a randomly distributed user. However, the waveguides and/or antennas in these four schemes still lack flexibility, which leads to a substantial performance gap compared with the proposed scheme. It is worth noting that, different from Fig. \ref{sg_L_rate}, the performance loss induced by imperfect CSI in Fig. \ref{sg_N_rate} is almost uniform for different values of $N$. This is because increasing $N$ adds more MWs/RF chains for beamforming, while the number of PAs combined within each MW remains unchanged. Therefore, the CSI error affects the same number of PA-to-user links in each MW, rather than causing an increasing accumulation of distance and phase deviations within each MW as in the case of increasing $L$ in Fig. \ref{sg_L_rate}. Moreover, the impact of CSI errors on the performance of the proposed algorithm for MW and/or PA position optimizations in the multi-user system will be further investigated in the next subsection.
\subsection{Multi-User System}
Next, we consider the multi-user scenario, i.e., $K > 1$. To comprehensively evaluate the different schemes, we consider the following user distributions: 
\begin{itemize}
	\item \textit{Horizontal (Hor.) user}: The users are randomly distributed within a narrow rectangular region, where their $x$-axis (horizontal) coordinates are uniformly distributed between 0 and $S_\mathrm{x}$, while their $y$-axis (vertical) coordinates are uniformly distributed in a narrow range between ${S_\mathrm{y}}/{2} - 0.05$ and $S_\mathrm{y}/2 +0.05$ m.
	\item \textit{Vertical (Ver.) user}: The users are randomly distributed within a narrow rectangular region, where their $y$-axis (vertical) coordinates are uniformly distributed between 0 and $S_\mathrm{y}$, while their $x$-axis (horizontal) coordinates are uniformly distributed in a narrow range between ${S_\mathrm{x}}/{2} - 0.05$ and $S_\mathrm{x}/2 +0.05$ m.
	\item \textit{Hotspot (Hot.) user}: Three non-overlapping circular hotspot areas with a radius of 0.5 m are randomly distributed within the rectangular region. The numbers of users in these three hotspots are 1, 2, and 3, respectively, and the users are randomly located within each hotspot area.
	\item \textit{Random (Ran.) user}: The users are randomly distributed within the entire rectangular region. 
\end{itemize}

\begin{figure*}[!t]
	\centering
	\subfloat[Horizontal and vertical user distributions. ]{\label{mu_P_hv}\includegraphics[width=1\columnwidth]{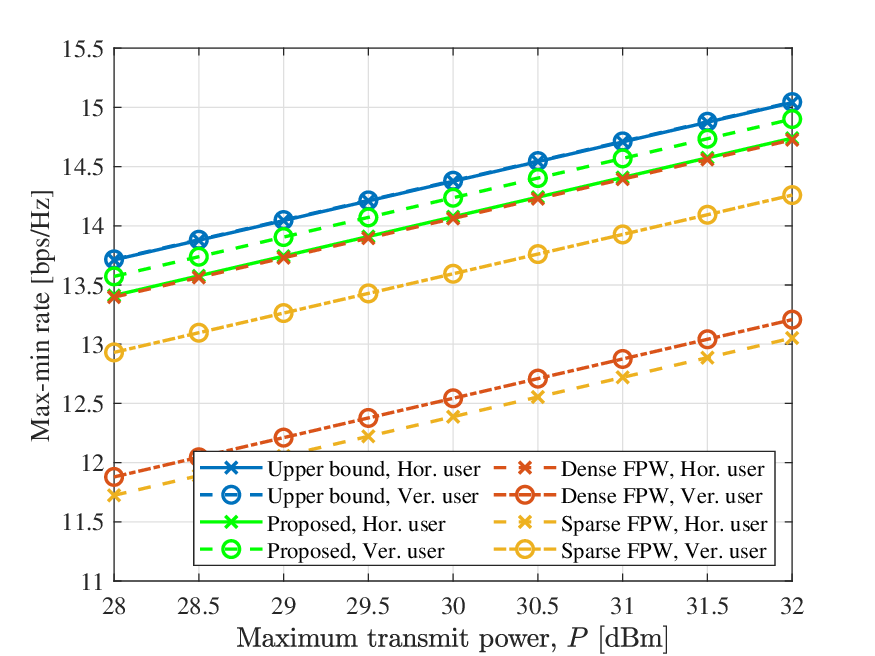}}
	\subfloat[Hotspot and random user distributions. ]{\label{mu_P_hr}\includegraphics[width=1\columnwidth]{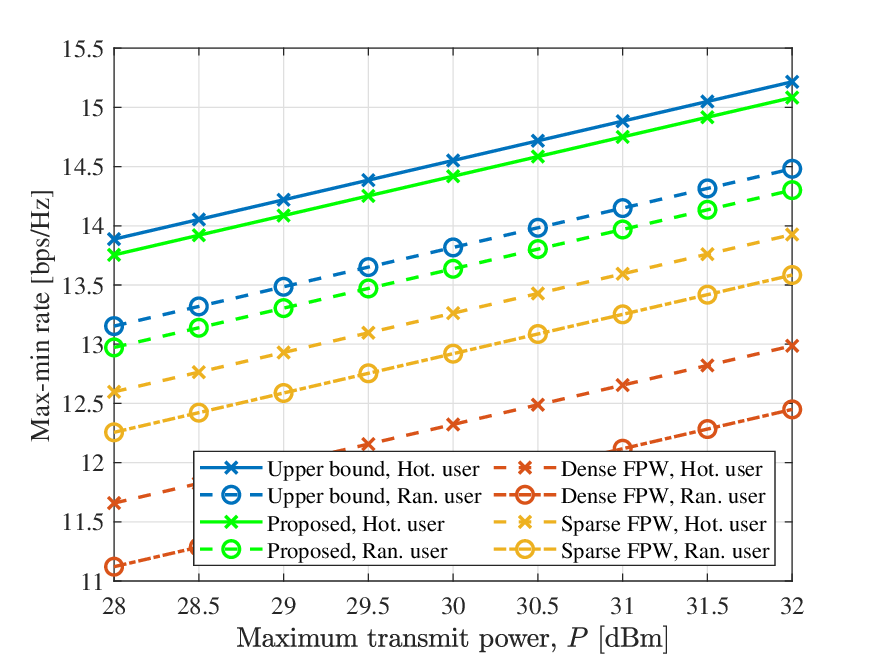}}
	\caption{Max-min rate versus maximum transmit power under different user distributions.}
	\label{mu_P}
\end{figure*}
\begin{figure*}[!t]
	\centering
	\subfloat[\textit{Proposed} scheme for horizontal users. ]{\label{MW_uhor}\includegraphics[width=0.66\columnwidth]{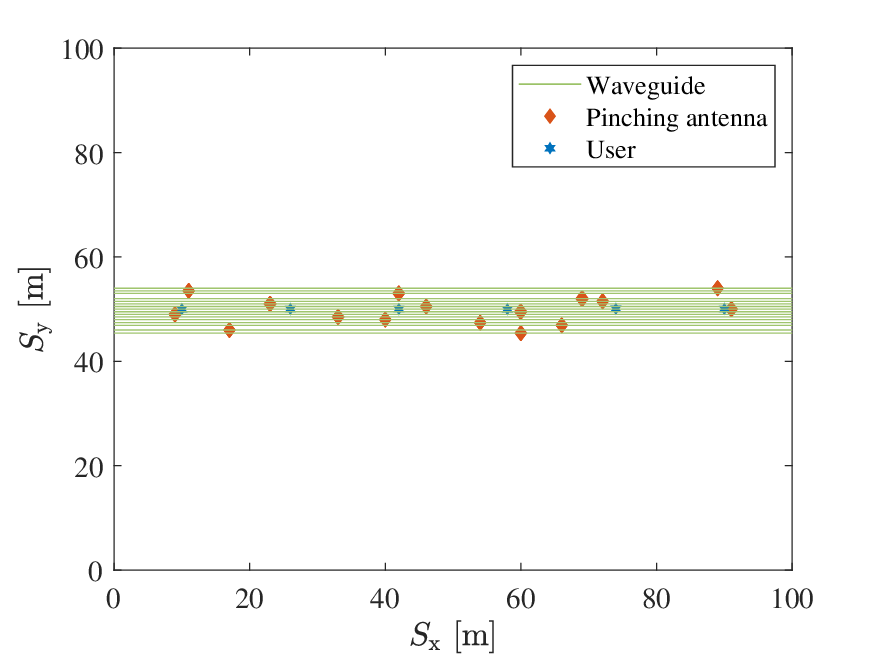}}
	\subfloat[\textit{Proposed} scheme for vertical users.]{\label{MW_uver}\includegraphics[width=0.66\columnwidth]{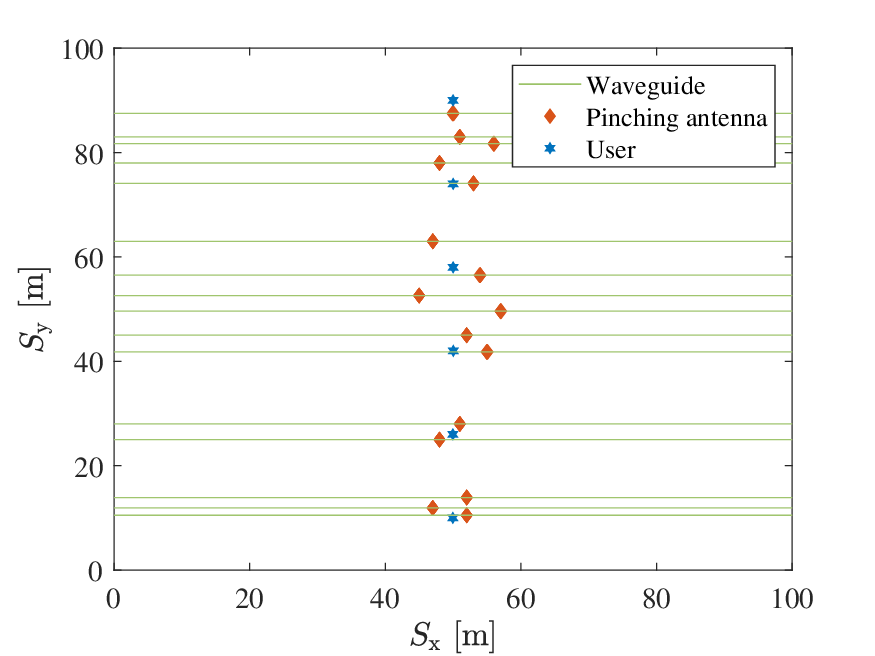}}
	\subfloat[\textit{Proposed} scheme for hotspot users.]{\label{MW_uhot}\includegraphics[width=0.66\columnwidth]{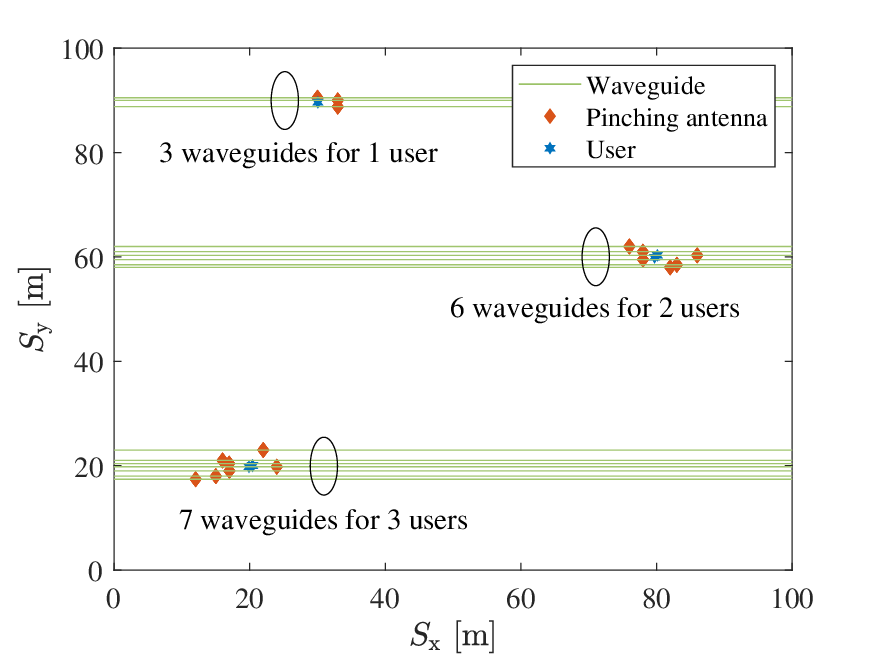}}\\
	\subfloat[\textit{Proposed} scheme for random users.]{\label{MW_uran}\includegraphics[width=0.66\columnwidth]{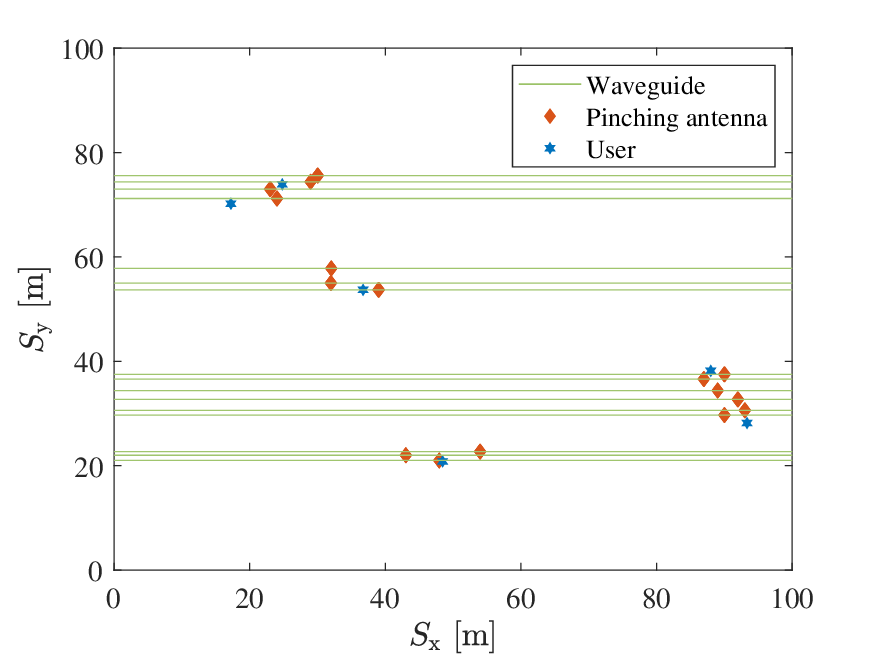}}
	\subfloat[\textit{Dense FPW} scheme for horizontal users.]{\label{FPWden_uhor}\includegraphics[width=0.66\columnwidth]{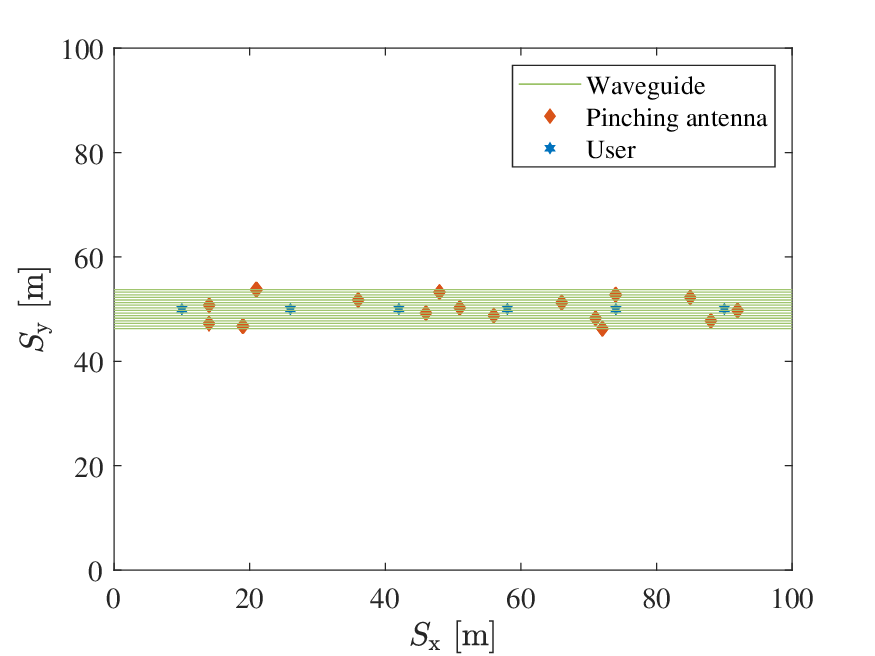}}
	\subfloat[\textit{Sparse FPW} scheme for vertical users.]{\label{FPWspa_uver}\includegraphics[width=0.66\columnwidth]{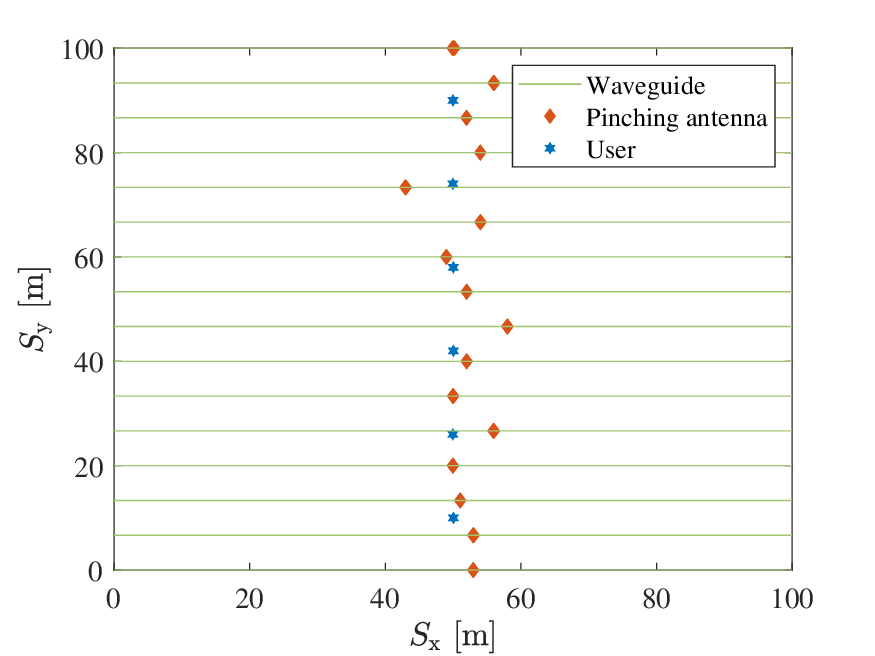}}
	\caption{Optimized MW and/or PA positions under different user distributions (top view).}
	\label{plot_pos}
\end{figure*}
Fig. \ref{mu_P} illustrates the max-min rate versus the maximum transmit power $P$ of the BS under different user distributions. In addition, Fig. \ref{plot_pos} presents one realization of the optimized MW and/or PA positions to provide an intuitive explanation for the performance differences among the considered schemes. We can see in Fig. \ref{mu_P}\subref{mu_P_hv} that the proposed scheme achieves favorable performance for both horizontal and vertical user distributions. This is because the mobility of MWs enables PASS to form antenna arrays with approximately horizontal or vertical structures in response to horizontally or vertically distributed users, as illustrated in Figs. \ref{plot_pos}\subref{MW_uhor} and \ref{plot_pos}\subref{MW_uver}, respectively. Unlike the proposed scheme, the dense FPW scheme deploys waveguides densely at the center of the rectangular region, which can form only a horizontal array to efficiently serve the horizontally distributed users located at the same center (see Fig. \ref{plot_pos}\subref{FPWden_uhor}), while being ineffective for vertically distributed users. Consequently, as shown in Fig. \ref{mu_P}\subref{mu_P_hv}, the dense FPW scheme achieves nearly the same performance as the proposed scheme for horizontal users; however, its performance degrades significantly for vertical users. In contrast, the sparse FPW scheme can exploit the sparse waveguide deployment to form vertical arrays via PA movement for the vertical user distribution (see Fig. \ref{plot_pos}\subref{FPWspa_uver}). However, its performance degrades for horizontally distributed users in Fig. \ref{mu_P}\subref{mu_P_hv}, since the PAs can only move horizontally along the waveguides. Furthermore, as shown in Fig. \ref{mu_P}\subref{mu_P_hr}, the proposed scheme achieves superior performance under the hotspot user distribution compared with the random user distribution, since it can allocate antenna resources on demand by jointly moving the MWs and PAs according to the locations of different hotspot areas and the user densities within them. For example, in Fig. \ref{plot_pos}\subref{MW_uhot}, three hotspot areas centered at (30 m, 90 m), (80 m, 60 m), and (20 m, 20 m) with the same size are served by 3, 6, and 7 waveguides to support 1, 2, and 3 users, respectively. Nevertheless, the proposed scheme in Fig. \ref{mu_P}\subref{mu_P_hr} still achieves significant performance gains over the two FPW schemes for the random user distribution, which can be attributed to the flexible MW deployment that more effectively adapts to the user locations (see Fig. \ref{plot_pos}\subref{MW_uran}). As a result, the MW introduces additional design flexibility to PASS, as the dynamic repositioning of MWs enables more effective adaptation to varying user distributions.

\begin{figure*}[!t]
	\centering
	\subfloat[Horizontal and vertical user distributions.]{\label{mu_S_hv}\includegraphics[width=1\columnwidth]{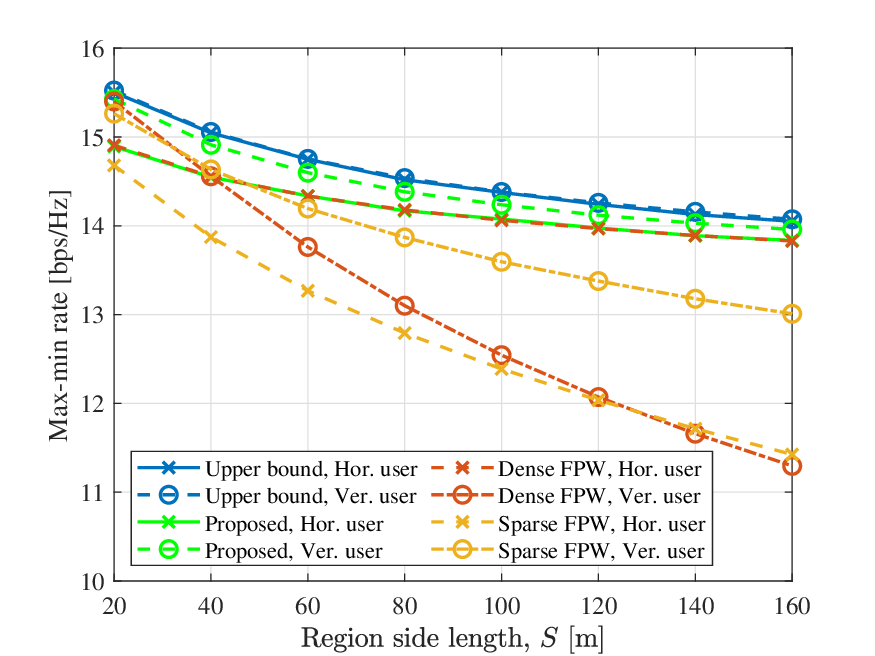}}
	\subfloat[Hotspot and random user distributions.]{\label{mu_S_hr}\includegraphics[width=1\columnwidth]{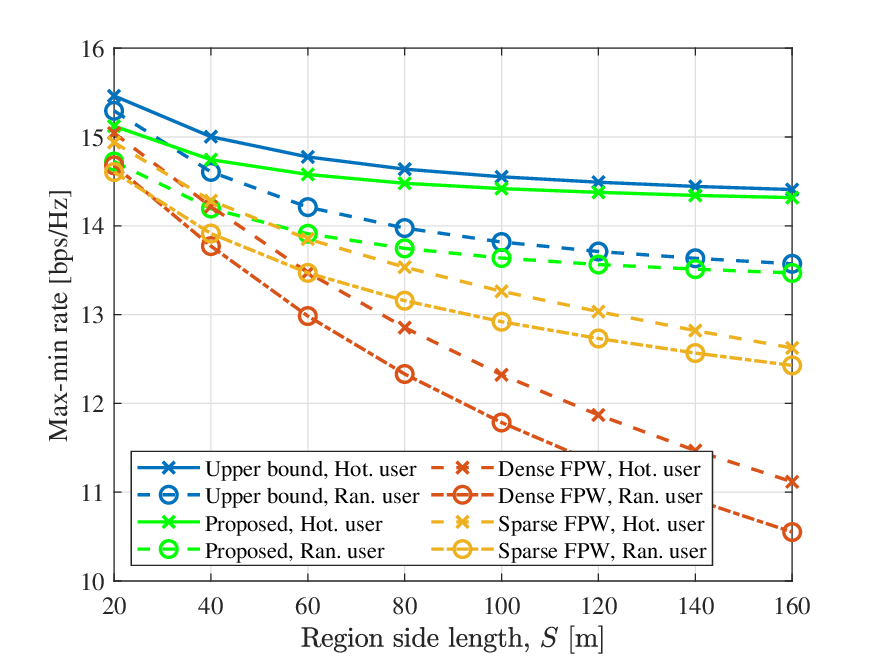}}
	\caption{Max-min rate versus region side length under different user distributions.}
	\label{mu_S}
\end{figure*}
Fig. \ref{mu_S} presents the impact of the user region size on the performance of different schemes. Without loss of generality, we assume that the rectangular region has sides of equal length, i.e., $S_\mathrm{x}=S_\mathrm{y}\triangleq S$. It can be seen that the max-min rates of all schemes decrease as $S$ increases due to the increased average distance between the antennas and the users. Moreover, the performance of the dense FPW scheme for vertical users and the sparse FPW scheme for horizontal users decreases more rapidly than that of the proposed scheme in Fig. \ref{mu_S}\subref{mu_S_hv}. This is because the fixed waveguide configurations in these two schemes restrict them to adapting only to either the horizontal or the vertical user distribution. Furthermore, for the random user distribution, the enlargement of the user region leads to a reduction in multi-user interference. Accordingly, Fig. \ref{mu_S}\subref{mu_S_hr} shows that as $S$ increases from 20 m to 160 m, the performance gap between the proposed scheme and its upper bound diminishes by 82\%. In addition, in the sparse FPW scheme, the uniform deployment of waveguides across the entire user region allows the antenna resources to adequately serve each hotspot or random user. This results in comparable performance as $S$ increases under both user distributions in Fig. \ref{mu_S}\subref{mu_S_hr}. However, since the sparse FPW scheme cannot flexibly adjust the positions of the waveguides, its performance is still inferior to the proposed scheme.

\begin{figure*}[!t]
	\centering
	\subfloat[Horizontal and vertical user distributions.]{\label{mu_M_hv}\includegraphics[width=1\columnwidth]{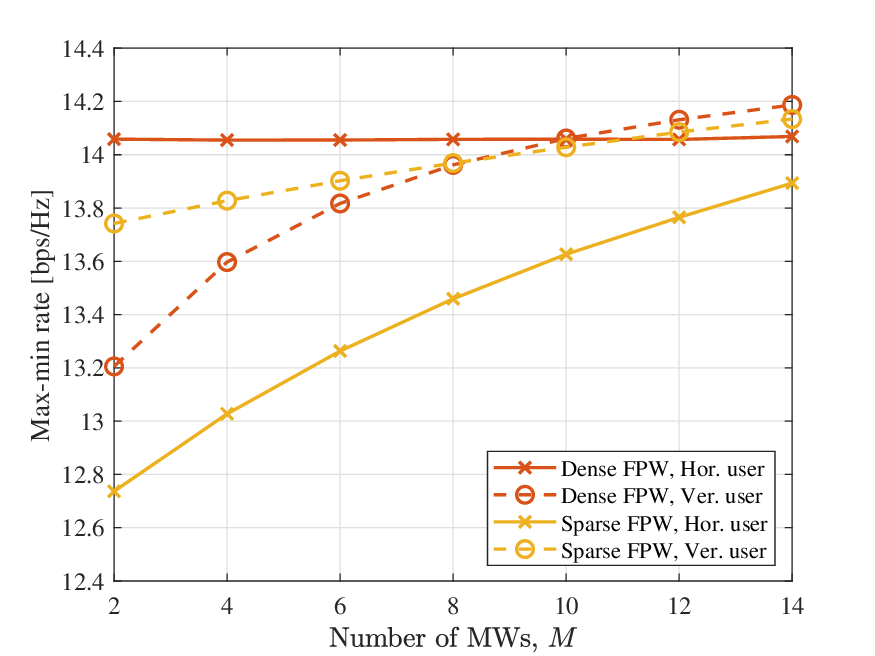}}
	\subfloat[Hotspot and random user distributions.]{\label{mu_M_hr}\includegraphics[width=1\columnwidth]{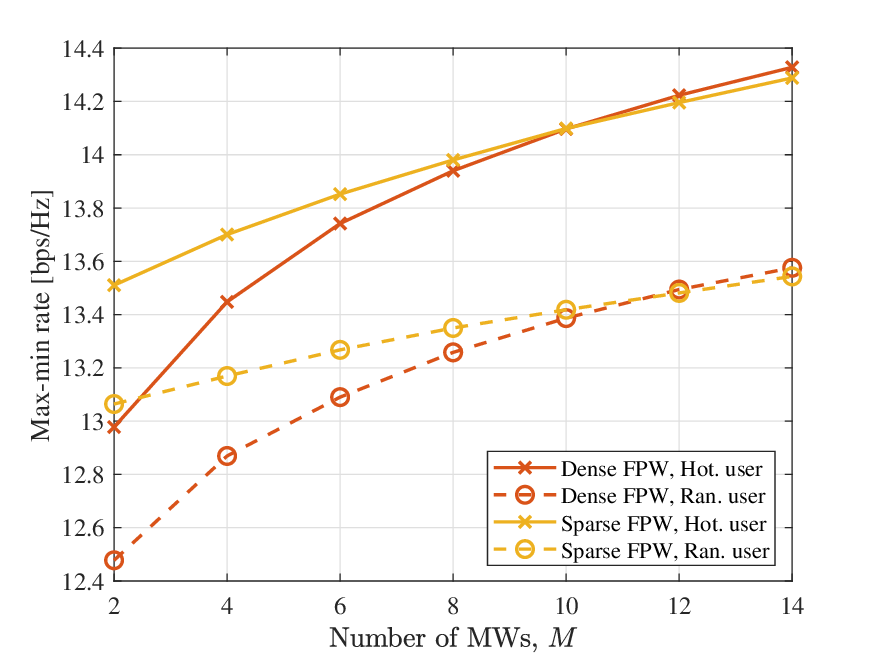}}
	\caption{Max-min rate versus number of MWs under different user distributions.}
	\label{mu_M}
\end{figure*}
To provide more insights for practical applications, Fig. \ref{mu_M} investigates the performance of the proposed scheme with partial waveguide mobility under different user distributions. Specifically, we assume that $M$ ($1 < M < N$) waveguides can be flexibly repositioned, while the remaining $N-M$ waveguides are fixed. These FPWs are either densely deployed based on \eqref{den} or sparsely deployed based on \eqref{spa}. As shown in Fig. \ref{mu_M}\subref{mu_M_hv}, when the FPWs are densely deployed, the performance remains unchanged for horizontally distributed users. The reason is as follows. For horizontally distributed users located in the middle of the user region, the waveguides in the dense FPW scheme are approximately positioned directly above the users, thereby naturally minimizing the distance between the antennas and the users, which is a special case of the proposed scheme. Moreover, the performance of all other schemes in Fig. \ref{mu_M}\subref{mu_M_hv} consistently increases with $M$ and eventually converges to the same level. The reason is that the increased number of MWs allows PASS to more effectively leverage the flexibility of waveguides to form different array geometries, thereby enhancing performance for both horizontal and vertical users. Besides, we can see in Fig. \ref{mu_M}\subref{mu_M_hr} that the performance gap of the proposed scheme between the hotspot-user case and the random-user case widens as $M$ increases. This is because the MWs can allocate antenna resources more efficiently according to the locations and densities of hotspot users, thereby improving spectral efficiency. Hence, the proposed scheme achieves more pronounced performance gains for hotspot users.

\begin{figure}[!t]
	\centering
	\includegraphics[width=1\linewidth]{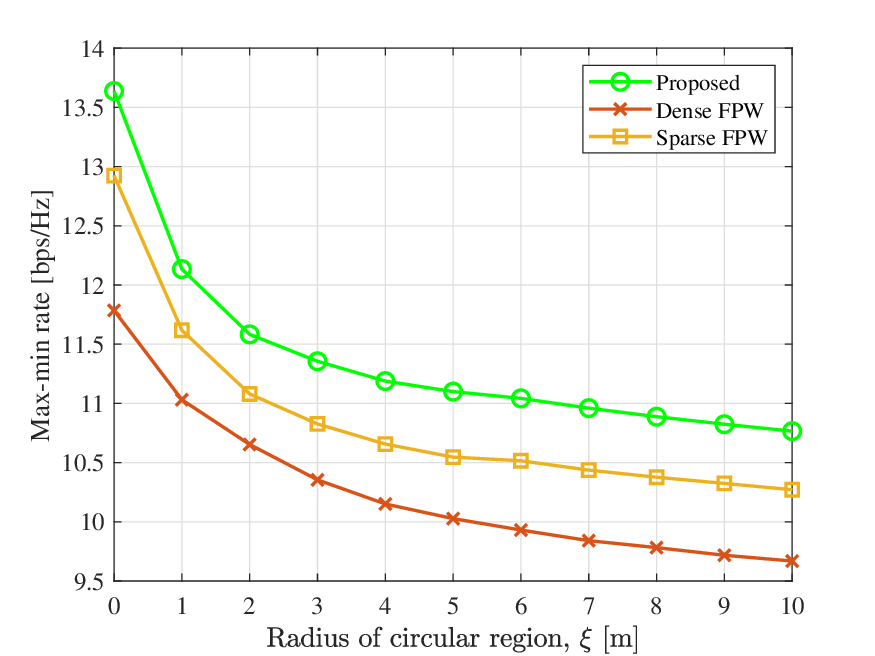}
	\caption{Impact of the CSI error on the performance of the proposed algorithm for MW and/or PA position optimizations.}
	\label{err}
\end{figure}
The above results are obtained under the assumption that the PASS has perfect CSI, which is specified by the user locations $\left\{\mathbf{u}_k^\mathrm{U}\right\}_{k=1}^K$. However, it is challenging to acquire accurate user locations in practice due to potential user mobility. Therefore, it is important to evaluate the impact of CSI errors induced by user mobility on the performance of the proposed algorithm for MW and/or PA position optimizations. To this end, we assume that each user randomly moves within a local region, where the actual user location is uniformly distributed inside a circular region centered at $\mathbf{u}_k^\mathrm{U}$ with radius $\xi$ m. Fig. \ref{err} depicts the max-min rate versus $\xi$ under the random-user case. To focus on the effect of imperfect CSI on MW and/or PA position optimizations, the MW and/or PA positions are optimized based on the estimated (imperfect) CSI, while the transmit beamforming vectors at the PASS are calculated using the actual channel vectors. As shown in Fig. \ref{err}, the max-min rate decreases slightly as $\xi$ increases because of the deviation between the estimated and actual channels. Nevertheless, the proposed scheme consistently maintains a significant performance advantage over the two FPW schemes because it jointly optimizes the MW positions along the $y$-axis and the PA positions along the $x$-axis. In contrast, the dense and sparse FPW schemes can only adjust the PA positions along fixed waveguides, and thus have limited flexibility to mitigate performance degradation induced by CSI errors.

\begin{figure}[!t]
	\centering
	\includegraphics[width=1\linewidth]{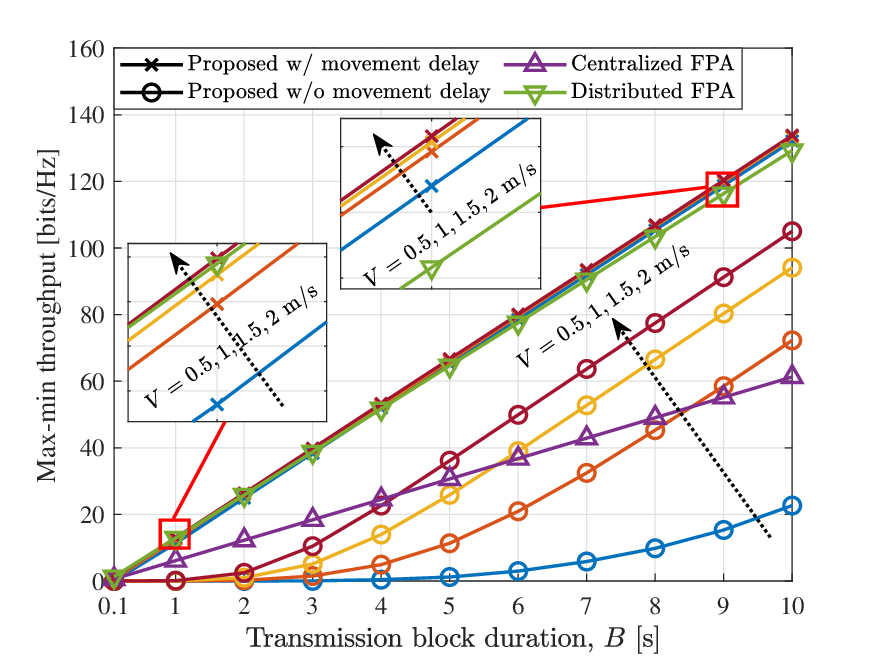}
	\caption{Max-min throughput versus transmission block duration for different MW moving speeds.}
	\label{delay}
\end{figure}
Fig. \ref{delay} evaluates the impact of MW movement delay on the max-min throughput over a service region of size $S_\mathrm{x}\times S_\mathrm{y}=160\mathrm{m}\times20\mathrm{m}$ with randomly distributed users, corresponding to practical deployment scenarios such as long indoor corridors or railway station platforms. In Fig. \ref{delay}, each transmission block with duration $B$ consists of an MW movement phase incurring a delay of $b$, followed by an information transmission phase. We only consider the mechanical movement delay of the MWs, while the PA positioning delay is ignored since the PA positions can be rapidly adjusted by electronically selecting the activated radiating points from densely pre-installed PAs along each waveguide \cite{PASS3}. Specifically, let $y_n^0$ and $y_n^\star$ denote the initial and target positions of the $n$th MW, respectively. The MW movement delay can be expressed as $b=\max_{1 \le n \le N}{\left|y_n^\star-y_n^0\right|}/{V}$, where $V$ denotes the MW moving speed. Accordingly, the max-min throughput over a transmission block (in bits/Hz) is given by $C=\left(B-b\right)^+\log_2\left(1+\Gamma \right) $, where $\Gamma$ denotes the same SNR of all users as defined in \eqref{sameSNR}. Based on the specifications of typical commercial motorized linear stages or belt-driven linear modules \cite{Zaber_linear_stage,Rollon_belt_actuator}, the MW moving speed is set to $V=0.5,1,1.5, 2$ m/s in the simulations. It is observed from Fig. \ref{delay} that simply maximizing the user rate without (w/o) considering movement delay may lead to poor throughput performance, especially when both $B$ and $V$ are small. This is because the no-delay optimization tends to move the MWs to positions that improve the instantaneous rates, but the resulting large movement distance substantially reduces the remaining time available for information transmission under finite transmission block durations. By contrast, the proposed scheme with (w/) movement delay consideration effectively balances the rate gain and the movement delay, and thus achieves higher max-min throughput, as shown in the right subfigure of Fig. \ref{delay}. However, when both $B$ and $V$ are small, the throughput performance of the proposed scheme is still worse than that of the distributed FPA scheme due to the non-negligible mechanical movement delay of the MWs, as shown in the left subfigure of Fig. \ref{delay}. These results indicate that frequent MW movement is not always desirable when the transmission block duration is comparable to the movement delay or when the channel coherence time is short. Therefore, for scenarios with stringent latency requirements, it is more suitable to adopt the two-timescale optimization scheme in Section \ref{two} to improve the long-term system performance.

\begin{figure}[!t]
	\centering
	\includegraphics[width=1\linewidth]{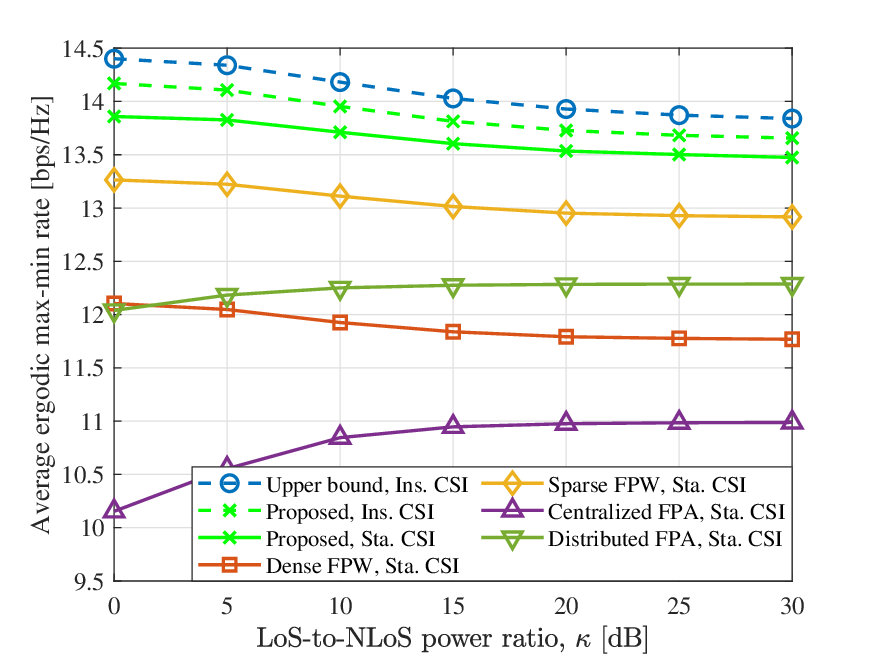}
	\caption{Average ergodic max-min rate versus LoS-to-NLoS power ratio for two-timescale optimization.}
	\label{kappa}
\end{figure}
To evaluate the effectiveness of the proposed two-timescale optimization framework and investigate the impact of multi-path propagation, Fig. \ref{kappa} plots the average ergodic max-min rates of various schemes versus the LoS-to-NLoS power ratio with randomly distributed users, where we assume that the LoS-to-NLoS power ratios of all users are identical, i.e., $\kappa_{k} \triangleq \kappa$, $\forall 1 \le k \le K$, and that the scatterers associated with each user are randomly distributed within a sphere of radius 5 m centered at the corresponding user. The ergodic max-min rate is calculated as the empirical expectation of $T=100$ instantaneous max-min rates, which is further averaged over 100 realizations of user distributions. For a comprehensive assessment under statistical (sta.) CSI, Fig. \ref{kappa} also includes the performance of the proposed scheme when instantaneous (ins.) CSI is available, together with the corresponding upper bound. As $\kappa$ increases, the rate performance of all schemes employing MW and/or PA position optimization degrades. This is because a larger $\kappa$ implies lower NLoS path power, which reduces the path diversity gain that can be exploited by MW/PA position optimization. By contrast, the rate performance of the two FPA schemes increases with $\kappa$, since the NLoS paths generally experience larger path loss than the LoS path, and thus a LoS-dominated propagation environment enables the transmitted signal power to be more efficiently received by the users. Nonetheless, the proposed scheme can still exploit the mobility of the MWs and PAs to maintain a clear advantage over benchmark schemes. In addition, as $\kappa$ increases from 0 dB to 30 dB, the performance gap between the proposed schemes with instantaneous CSI and statistical CSI is reduced by 41\%, which indicates the superior effectiveness of the proposed MW-enabled PASS scheme when fewer dominant channel paths are available, as commonly encountered in high-frequency communication systems.

\begin{table}[!t]
	\centering
	\caption{Comparisons of Complexity and Performance}
	\label{tab2}
	\renewcommand{\arraystretch}{1.4}
	\begin{tabular}{|c|c|c|c|}
		\hline
		\textbf{Scenario} & \textbf{Scheme} & \textbf{Run time [s]} & \makecell{\textbf{Achievable}\\\textbf{rate [bps/Hz]}} \\
		\hline
		\multirow{3}{*}{Single-user} 
		& Proposed   & 7.41  & 18.56 \\
		\cline{2-4}
		& Dense FPW  & 9.48  & 16.02 \\
		\cline{2-4}
		& Sparse FPW & 8.13  & 16.56 \\
		\hline
		\multirow{3}{*}{Multi-user}
		& Proposed   & 26.32 & 13.47 \\
		\cline{2-4}
		& Dense FPW  & 18.23 & 11.62 \\
		\cline{2-4}
		& Sparse FPW & 21.48 & 12.75 \\
		\hline
	\end{tabular}
\end{table}

Finally, Table \ref{tab2} compares the computational complexity\footnote{The simulations are implemented in MATLAB R2023a on a 64-bit Windows platform equipped with an Intel i7-12700F central processing unit (CPU).} and performance of the proposed scheme and the two FPW schemes under both the single-user scenario and the multi-user scenario with randomly distributed users. In the single-user case, the proposed scheme provides a closed-form optimal solution for the MW positions, which accelerates the convergence of the PA position optimization. As a result, it achieves lower run time and higher achievable rate than the two FPW schemes. In the multi-user case, the proposed scheme achieves superior performance by jointly optimizing the MW and PA positions, at the cost of increased computational complexity. Therefore, the partial-waveguide-mobility design in Fig. \ref{mu_M} and the two-timescale optimization framework in Fig. \ref{kappa} can be adopted to better balance computational efficiency and system performance.
\section{Conclusion}\label{6}
This paper proposed a new MW architecture, where each MW is connected to an RF chain via a flexible cable and can be moved linearly with the aid of drivers such as motors. By dynamically adjusting the MW and PA positions, PASS can adaptively track user locations and form favorable array geometries tailored to different user distribution characteristics. We first considered the single-user scenario, where the closed-form solutions for the optimal MW positions and an upper bound on the user rate were derived. In addition, for the case of multiple PAs on each MW, we proposed a low-complexity 1D search algorithm to optimize the PA positions. Then, for the multi-user scenario, we derived the upper bounds on the minimum rate among all users and analyzed the conditions required to achieve these performance bounds. To maximize the minimum rate of all users, we proposed a two-scale search algorithm, where the large-scale search coarsely determines the MW and PA positions and the small-scale search finely tunes them, thereby efficiently optimizing MW and PA positions. The joint design of the MW and PA positions was further extended to the case with statistical CSI, where the MW positions are updated on a large timescale to reduce the mechanical movement overhead of MWs, while the PA positions and beamforming are optimized on a small timescale based on instantaneous CSI. Simulation results demonstrated that PASS with MWs can enhance the flexibility of conventional PASS with FPWs, thereby adapting more effectively to different user distributions. These results provided valuable insights for the practical implementation of MW-enabled PASS in future wireless communication networks.

\begin{IEEEbiography}[{\includegraphics[width=1in,height=1.25in,clip,keepaspectratio]{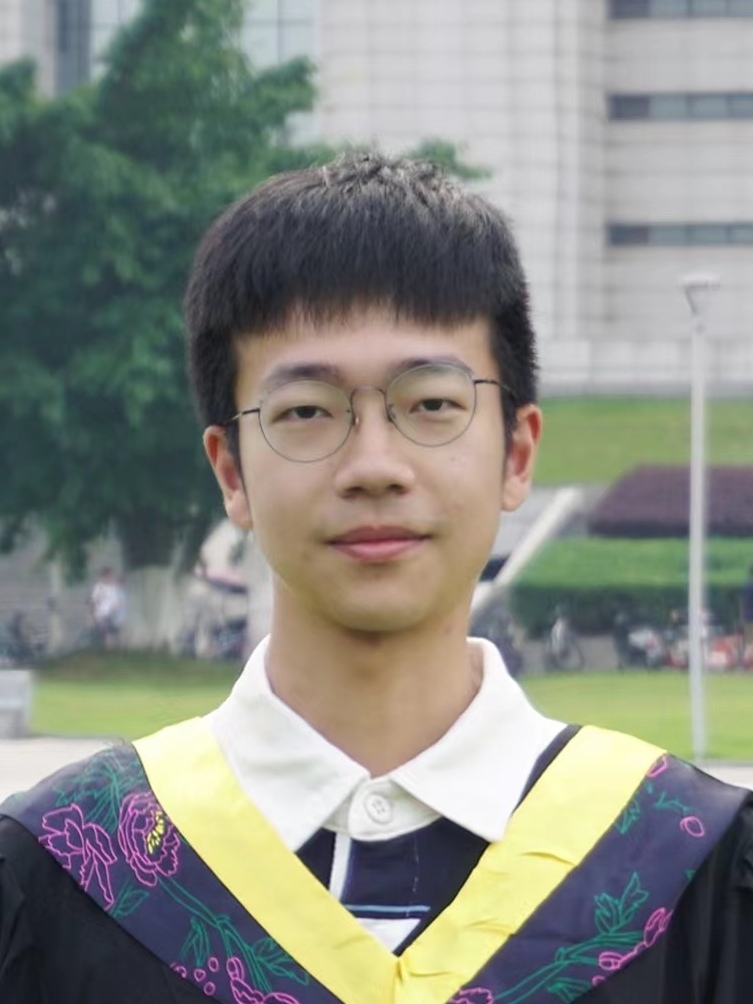}}]
	{Jingze Ding}(Graduate Student Member, IEEE) received the B.E. degree in information engineering from the University of Electronic Science and Technology of China (UESTC), Chengdu, China, in 2022. He is currently pursuing the Ph.D. degree with the School of Electronics, Peking University, Beijing, China. He is also a visiting Ph.D. student with the Department of Electrical and Computer Engineering, National University of Singapore, Singapore. His research interests include flexible antenna and optimization techniques.
\end{IEEEbiography}

\begin{IEEEbiography}[{\includegraphics[width=1in,height=1.25in,clip,keepaspectratio]{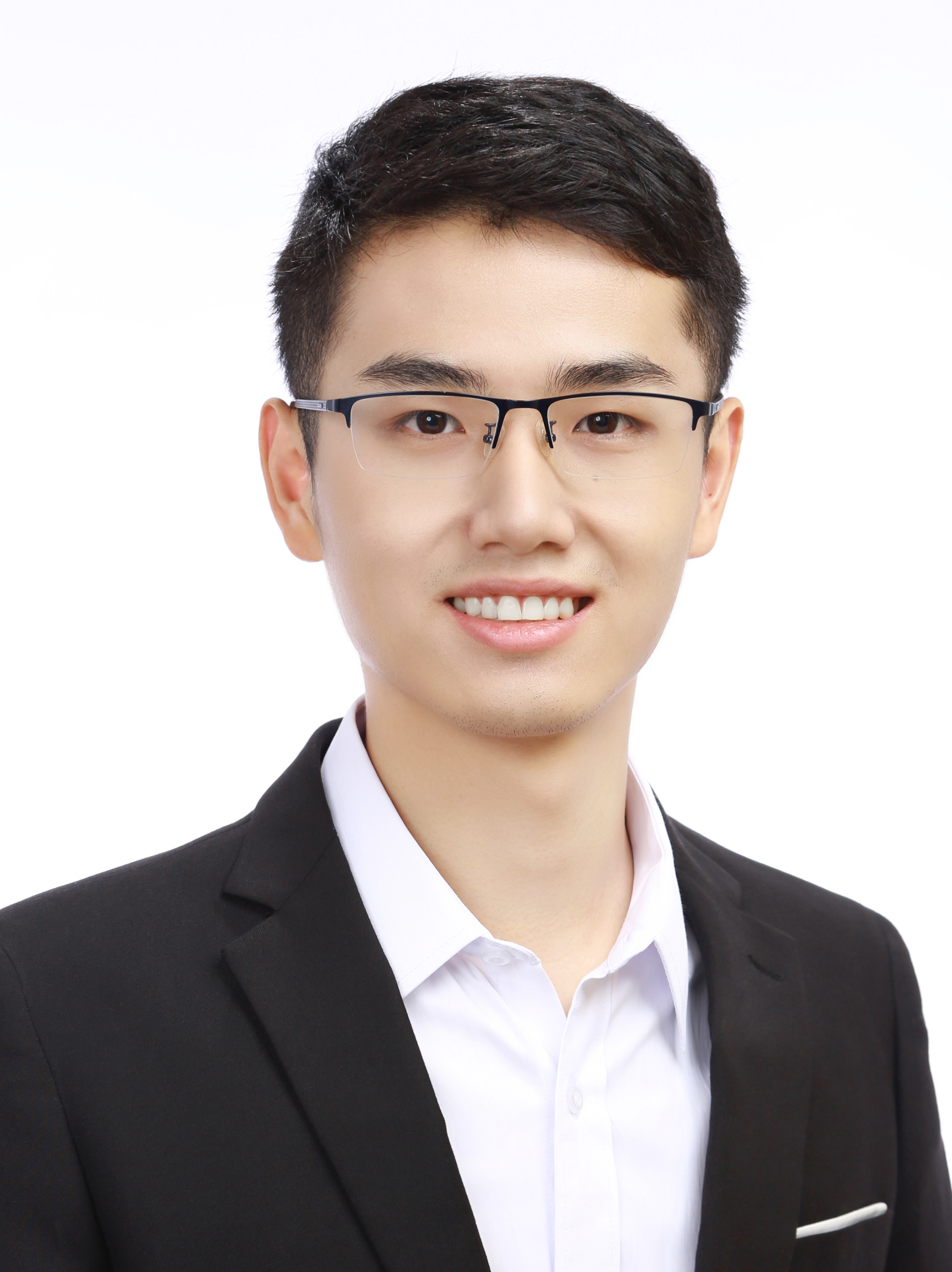}}]
	{Zijian Zhou}(Member, IEEE) received the B.Eng. degree in information engineering from Soochow University, Suzhou, China, in 2020. He is currently pursuing the Ph.D. degree with the School of Science and Engineering, The Chinese University of Hong Kong, Shenzhen, China. From 2020 to 2024, he was a Research Assistant with the School of Electronics, Peking University, Beijing, China. His research interests include polarforming for wireless communications, movable antenna (MA)-enabled communication systems, and convex optimization. He was honored as a Chun-Tsung scholar of Soochow University in 2019.
\end{IEEEbiography}

\begin{IEEEbiography}[{\includegraphics[width=1in,height=1.25in,clip,keepaspectratio]{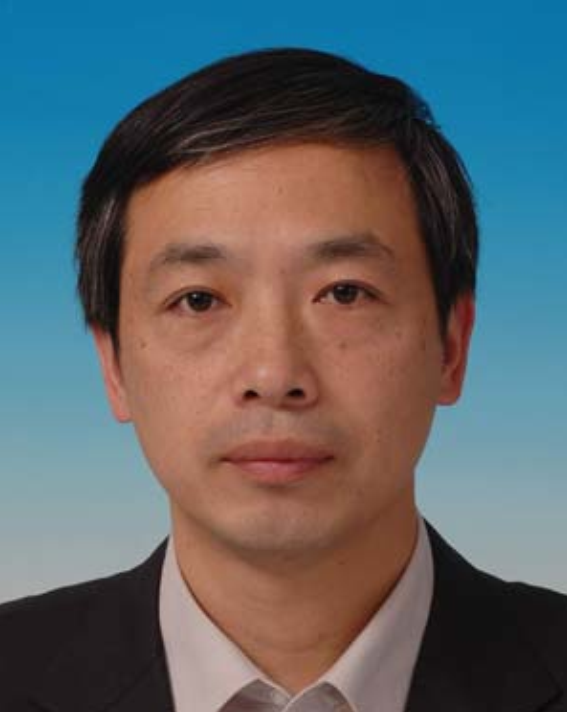}}]
	{Bingli Jiao}(Senior Member, IEEE) received the B.S. and M.S. degrees from Peking University, Beijing, China, in 1983 and 1988, respectively, and the Ph.D. degree from Saarland University, Saarbrcken, Germany, in 1995. He became an Associate Professor in 1995 and a Professor with Peking University in 2000. He is currently the Director of the Joint Laboratory for Advanced Communication Research between Peking University and Geespace. He is a pioneer of co-frequency and co-time full-duplex as found in his early patent in 2006. His research interests include full-duplex communications, information theory, and signal processing.
\end{IEEEbiography}

\begin{IEEEbiography}[{\includegraphics[width=1in,height=1.25in,clip,keepaspectratio]{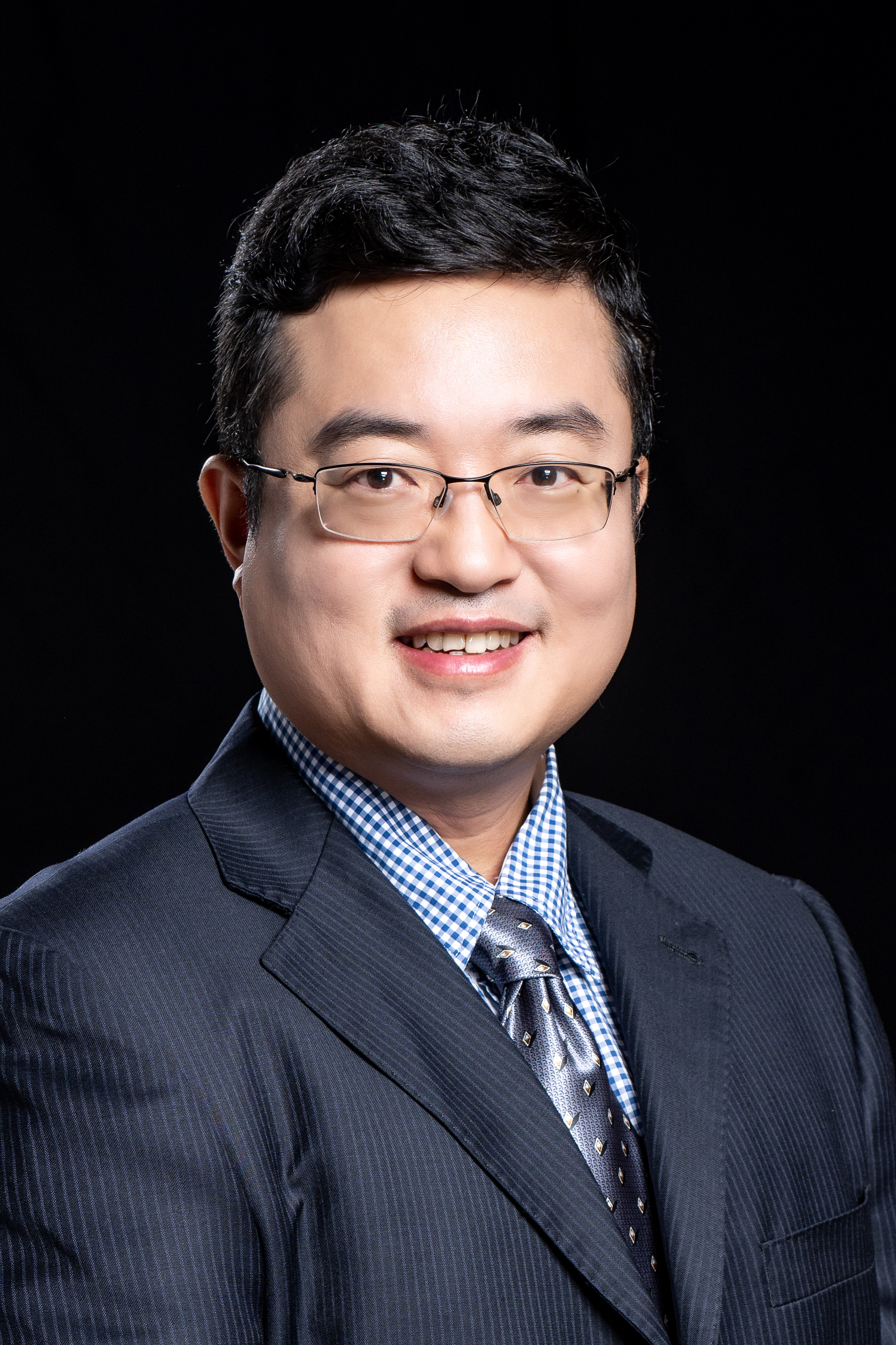}}]
	{Rui Zhang}(Fellow, IEEE) received the B.Eng. (first-class Hons.) and M.Eng. degrees from the National University of Singapore, Singapore, and the Ph.D. degree from the Stanford University, Stanford, CA, USA, all in electrical engineering.
	
	From 2007 to 2009, he worked as a research scientist at the Institute for Infocomm Research, ASTAR, Singapore. In 2010, he joined the Department of Electrical and Computer Engineering of National University of Singapore, where he is now a Provost's Chair Professor. He is also an Adjunct Professor with the School of Science and Engineering, The Chinese University of Hong Kong, Shenzhen, China. He has published over 600 papers, all in the field of wireless communications and networks. He has been listed as a Highly Cited Researcher by Thomson Reuters/Clarivate Analytics since 2015. His current research interests include intelligent surfaces, reconfigurable antennas, radio mapping, non-terrestrial communications, wireless power transfer, AI and optimization methods.      
	
	He was the recipient of the 6th IEEE Communications Society Asia-Pacific Region Best Young Researcher Award in 2011, the Young Researcher Award of National University of Singapore in 2015, the Wireless Communications Technical Committee Recognition Award in 2020, the IEEE Signal Processing and Computing for Communications (SPCC) Technical Recognition Award in 2021, and the IEEE Communications Society Technical Committee on Cognitive Networks (TCCN) Recognition Award in 2023. His works received 18 IEEE Best Journal Paper Awards, including the IEEE Marconi Prize Paper Award in Wireless Communications in 2015 and 2020, the IEEE Signal Processing Society Best Paper Award in 2016, the IEEE Communications Society Heinrich Hertz Prize Paper Award in 2017, 2020 and 2022, the IEEE Communications Society Stephen O. Rice Prize in 2021, etc. He served for over 30 international conferences as the TPC co-chair or an organizing committee member. He was an elected member of the IEEE Signal Processing Society SPCOM Technical Committee from 2012 to 2017 and SAM Technical Committee from 2013 to 2015. He served as the Vice Chair of the IEEE Communications Society Asia-Pacific Board Technical Affairs Committee from 2014 to 2015, a member of the Steering Committee of the IEEE Wireless Communications Letters from 2018 to 2021, a member of the IEEE Communications Society Wireless Communications Technical Committee (WTC) Award Committee from 2023 to 2025. He was a Distinguished Lecturer of IEEE Signal Processing Society and IEEE Communications Society from 2019 to 2020. He served as an Editor for several IEEE journals, including the IEEE TRANSACTIONS ON WIRELESS COMMUNICATIONS from 2012 to 2016, the IEEE JOURNAL ON SELECTED AREAS IN COMMUNICATIONS: Green Communications and Networking Series from 2015 to 2016, the IEEE TRANSACTIONS ON SIGNAL PROCESSING from 2013 to 2017, the IEEE TRANSACTIONS ON GREEN COMMUNICATIONS AND NETWORKING from 2016 to 2020, and the IEEE TRANSACTIONS ON COMMUNICATIONS from 2017 to 2022. He now serves as an Editorial Board Member of npj Wireless Technology, and the Chair of the IEEE Communications Society Wireless Communications Technical Committee (WTC) Award Committee. He is a Fellow of the Academy of Engineering Singapore.
\end{IEEEbiography}
\end{document}